\documentclass[vecphys]{svmult}

\usepackage{makeidx}         
\usepackage{graphicx}        
\usepackage{multicol}        
\usepackage{natbib}        
\usepackage{journals}        
\usepackage[bottom]{footmisc}

\makeindex             

\begin{document}

\title*{On the Formation and Dynamical Evolution of Planets in Binaries}
\titlerunning{Formation and Dynamical Evolution of Planets in Binaries}
\author{
Willy Kley \inst{1}
\and
Richard P. Nelson \inst{2}
}
\institute{
     Institut f\"ur Astronomie \& Astrophysik,
     Universit\"at T\"ubingen, \\
     Auf der Morgenstelle 10, D-72076 T\"ubingen, Germany,
\texttt{wilhelm.kley@uni-tuebingen.de}
\and
     Astronomy Unit,
     Queen Mary, University of London,
     Mile End Road,
     London E1 4NS, United Kingdom,
\texttt{r.p.nelson@qmul.ac.uk}
}

%
%
\maketitle

\abstract{
Among the extrasolar planetary systems about 30 are located in a stellar binary
orbiting one of the stars, preferably the more massive primary. The dynamical influence
of the second companion alters firstly the orbital elements of the forming
protoplanet directly and secondly the structure of the disk from which the planet formed
which in turn will modify the planet's evolution.
We present detailed analysis of these effects and present new hydrodynamical simulations
of the evolution of protoplanets embedded in circumstellar disks in the presence
of a companion star, and compare our results to the system $\gamma$~Cep.
To analyse the early formation of planetary embryos, we follow the evolution of a swarm
of planetesimals embedded in a circumstellar disk. Finally, we study the evolution of 
planets embedded in circumbinary disks.
}

\section{Introduction}
\label{sec:intro}

\subsection{Summary of observations} 

At the time of writing, approximately 29 extrasolar planets have been
discovered in binary star systems, all of which are orbiting about
a single component of the binary. For a review of the global statistics
see the papers by \citet{2004A&A...417..353E} and \citet{2007astro.ph..3795M},
as well as the relevant chapters in this book (Eggenberger \& Udry).
So far, there have been no discoveries of circumbinary planets.
The binary star systems that host planets are very diverse
in their properties, with binary semimajor axes ranging from $\simeq 6400$ AU
down to $\simeq 20$ AU. In the case where the orbits are eccentric,
the binary periastron can be as small as $\simeq 12$ AU, such that
important dynamical effects are expected to have occured during
and after planet formation. One such example is the well studied
system $\gamma$ Cep \citep{2003ApJ...599.1383H} which contains a planet
of mass $ m \sin{i} \simeq 2$ Jupiter masses 
with a semimajor axis of $\simeq 2$ AU.
Here the binary semimajor axis is $\simeq 18$ AU 
and periastron is $\simeq 12$ AU.
Another interesting case is GL86 \citep{2005MNRAS.361L..15M},
which consists of a binary system whose secondary is a $\simeq 0.55$ M$_{odot}$
white dwarf whose projected orbital separation is $\simeq 21$ AU. 
GL86 is reported to host a planet with $m \, sin{i} \simeq 4$ Jupiter masses
\citep{2000A&A...354...99Q}. It is worth noting that the white
dwarf projenitor was probably a Solar mass main sequence star, such that
the orbital separation was even smaller in the past.

Clearly the close binary systems containing planets provide an excellent
laboratory for testing theories of planet formation, as the
presence of the companion may create conditions normally thought
to be inconducive to planet formation. It is these closer systems
that we mainly focus on in this article.

\subsection{Summary of planet formation in binaries}  
\label{sec:formation}
In a binary star system the early formation of planets may be strongly
influenced by 
changes in the structure of the protoplanetary disk caused by tidal
forces from the binary companion. For a circumstellar 
disk, significant effects will occur if the disk outer edge is tidally
truncated by the binary companion, as strong spiral shock waves will
be lauched near the disk edge and propagate inward. For a 
circumstellar disk in a binary system which is not subject to
strong tidal forcing, it seems likely that the effect of the
companion star will be modest, unless the orbital inclinations are such
that the Kozai effect becomes important \citep{1997AJ....113.1915I}.
In a circumbinary disk one can
almost always expect strong tidal interaction between the binary and
disk, and hence significant effects on planet formation. 
In this article we restrict our discussion to two basic scenarios.
The first is planet formation and evolution in a circumstellar disk
around the primary (most massive) star - although we note that of
the 29 binary systems with known planets, two of them 
host planets around the secondary star (16 Cyg and HD178911).
The second scenario is planet formation in circumbinary disks. We restrict
our discussion to those early phases of planetary formation that occur
in a gas rich environment.

In a circumstellar disk, the tidal torques of the companion 
star generate strong spiral shocks,
and angular momentum is transferred
to the binary orbit. This in turn leads to disk truncation.
Using analytical and numerical methods \citet{1994ApJ...421..651A} show how the
truncation radius $r_t$ of the disk depends on the 
binary semimajor axis $a_{bin}$, its
eccentricity $e_{bin}$, the mass ratio $q = M_2/M_1$ (where $M_1$, $M_2$ denote
are the primary and secondary mass, respectively), 
and the viscosity $\nu$ of the disk.
For typical values of $q \approx 0.5$ and $e_{bin}=0.3$ the disk will be truncated
to a radius of $r_t \approx 1/3 a_{bin}$ for 
typical disk Reynold's numbers of $10^5$
\citep{1994ApJ...421..651A, 1996MNRAS.282..597L, 1999MNRAS.304..425A}.
For a given mass ratio $q$ and semi-major axis $a_{bin}$ an increase in $e_{bin}$ will
reduce the size of the disk while a large $\nu$ will increase the disks
radius.
Not only will the disk be truncated, but
the overall structure may be modified by the binary companion. In 
section \ref{sec:circumstellar}
we will illustrate this effect.

In a circumbinary disk, the binary creates a tidally-induced inner
cavity. For typical disk and binary parameters (e.g. $e_{bin}=0.3$, $q=0.5$) 
the size of the cavity is $\simeq 2.7 \times a_{bin}$
\citep{1994ApJ...421..651A}.

Whether these changes in the disk structure in circumstellar of circumbinary
systems have an influence on the likelihood
of planet formation in such disks has long been a matter of debate.
The dynamical action of the binary has several potential consequences which
may be adverse to planet formation: 
{\it i}) it changes the stability properties of orbits,
{\it ii}) it increases the velocity dispersion of planetesimals
{\it iii}) it reduces the life--time of the disk, and 
{\it iv}) it increases the temperature in the disk.

In a numerical study \citet{2000ApJ...537L..65N} investigated the evolution of 
an equal mass binary with a 50~AU separation and an eccentricity of $0.3$. 
He argued that both main scenarios of giant
planet formation (i.e. through core instability or gravitational instability)
are 
strongly handicapped, because the eccentric companion will induce a periodic heating
of the disk up to temperatures possibly above 1200 K. Since the condensation
of particles as well as the occurence of gravitational instability require
lower temperatures, planet formation will be made more difficult.
Clearly the strength of this effect will depend on the binary separation
and its mass ratio.
In addition to the approach taken by \citet{2000ApJ...537L..65N} 
the influence a stellar companion has on the evolution of a massive planet
embedded in a circumstellar disk has been investigated by
\citet{2000IAUS..200P.211K}, where the evolution of the embedded planet 
has been studied through hydrodynamical simulations 
\citep[see also the review article by][]{2000ASPC..219..189K}.
However, in these preliminary simulations only very short time spans have been
covered and the initial disk configuration may have been unrealistic.

Recent numerical studies of the final stages of terrestrial planet
formation in rather close binaries with separations of only 20--30 AU, 
that involve giant impacts between $\sim$ lunar mass 
planetary embryos, show that
it is indeed possible to form terrestrial planets in such systems
\citep{2004RMxAC..22...99L, 2005MSAIS...6..172T, 2007astro.ph..1266Q},
provided it is possible for the planetary embryos themselves to form.

It is already the case for planet formation around single stars that
the life--time of the
disk represents a limiting factor in the formation of planets.
It has been suspected that the dynamical action of a companion will reduce
the life--time of disks substantially.
However, a recent analysis of the observational data of disks in binary stars
finds no or very little change in the lifetimes of the disks,
at least for separations larger than about 20 AU \citep{2007prpl.conf..395M}.

The early phase of planetesimal formation and subsequent formation of
Earth-like planets is described in more detail in other chapters of this book.
Here we will concentrate on the formation and evolution of planets
in a gas rich environment, where inclusion of the full dynamics of the
protoplanetary disk is crucial. We consider the dynamics
of planetesimals, low mass planets, and high mass planets in circumstellar
and circumbinary disks.

\section{Evolution of planets in circumstellar disks with a companion} 
\label{sec:circumstellar}
The presence of a companion star influences the structure of a circumstellar
disk around the primary star due to gravitational torques acting on the disk.
This leads to an exchange of energy and angular momentum between the binary and
the disk.
For close binaries the disk becomes truncated where the truncation
radius $r_t$ depends primarily on the parameters of the binary,
i.e. the mas ratio $q$, the semi-major axis $a_{bin}$ and eccentricity $e_{bin}$, and the
viscosity of the disk.
The radius $r_t$ has been calculated semi-analytically and numerically
by \citet{1994ApJ...421..651A}.

The effects of the companion on planet formation are likely to be 
most pronounced in binaries with separations $\le  20$ AU,
rather than in long period systems with $a_{bin} > 1000$ AU.
Among the very close binary stars containing planets is the well studied system
$\gamma$-Cep. Including observations taken over decades,
\citet{2003ApJ...599.1383H}
confirmed the presence of a planet orbiting the primary star in this system.
Very recently, new radial velocity measurements and additional Hipparcos data
have refined the binary orbit \citep{2007ApJ...654.1095T} and the direct
imaging of the secondary has fixed the masses of the binary to $M_1=1.4$ and $M_2=0.4 M_\odot$
\citep{2007A&A...462..777N}.
This system with a binary separation of about $20$~AU contains a massive planet
with a minimum mass of 1.6$M_{Jup}$ orbiting the primary star at a distance
of approximately 2.1 AU.
Assuming that the planet has not been captured at a later time,
or that the binary orbit has not shrunk since planet formation, this system
represents a very challenging environment for the formation of planets indeed,
and we choose it to illustrate the main influence a close companion has on the planet 
formation process.

\subsection{Disk evolution in the presence of a companion} 
When studying the formation of planets in a protoplanetary disk in the
presence of a secondary star it is necessary to first follow the evolution of the
perturbed disk without an embedded planet and bring the system 
into equilibrium,
before adding a planetary embryo at a later time.
 
We choose to model a specific system where the orbital elements of the
binary have been chosen to match the system $\gamma$ Cep quite closely.
The data for this system have been taken from \citep{2003ApJ...599.1383H} 
which do not include the most
recent improvements mentioned above \citep{2007A&A...462..777N}.
These newest refinements primarily concern the mass of the
primary and do not alter our conclusions at all. 
We are interested here in demonstrating
the principle physical effects rather than trying to 
achieve a perfect match with
all the observations of this particular system.

For this study we choose a binary with $M_1 = 1.59 M_\odot$, $M_2 = 0.38 M_\odot$,
$a_{bin} = 18.5$~AU and $e_{bin} = 0.36$, which translates into a binary period
of $P = 56.7$~yrs. We assume that the primary 
star is surrounded by a flat circumstellar disk, where
the binary orbit and the disk are coplanar.
In a numerical hydrodynamical model of the system, the fact that the disk's
vertical thickness $H(r)$ at a given distance $r$ from the primary
is typically small with respect to the radius ($H/r <<1$) is typically used to
perform restricted two-dimensional (2D) simulations and neglect the vertical
extent altogether. Here, we present such 2D hydrodynamical
simulations of a circumstellar disk which is perturbed by the secondary. 
We assume that the effects of the intrinsic turbulence of the disk
can be described approximately through the viscous Navier-Stokes equations,
which are solved by a finite volume method which is second order in
space and time. To substantiate our results we utilize two different codes
{\tt RH2D} \citep{1999MNRAS.303..696K, 1989A&A...208...98K} and 
{\tt NIRVANA} \citep{2000MNRAS.318...18N, 1997ZiegYork}.

\medskip

\noindent
{\it Numerical Setup}: 
As the disk is orbiting only one star we utilize an adapted cylindrical
coordinate system ($r$, $\varphi$) which is centered on the primary.
It extends radially from $r_{min} = 0.5$~AU to $r_{max}=8$~AU and in
azimuth around a whole annulus ($\varphi_{min} = 0, \varphi_{max} = 2 \pi$).
Within this domain at the beginning of the simulations ($t=0$)
an axisymmetric disk (with respect to the primary) is initialized with a
surface density profile $\Sigma(r) = \Sigma_0 r^{-1/2}$ where the reference density
$\Sigma_0$ is chosen such that the total mass in the compuational domain
(within $r_{min}$ and  $r_{max}$) equals $1.75 \cdot 10^{-3} M_\odot$ which
implies $\Sigma_0 = 1.89 \cdot 10^{-5} M_{sol}$/AU$^2$.
The temperature profile is fixed here and given by $T(r) \propto r^{-1}$ which
follows from the assumed constancy of $h = H/r$ which is fixed to $h =0.05$.
For the viscosity we assume an $\alpha$-type prescription where the coefficient of the
kinematic viscosity is given by $\nu = \alpha c_s H$ with $\alpha = 0.005$, and
the sound speed $c_s(r) = h \, v_{kep}(r)$.

The boundary conditions are chosen such that material may escape through the radial
boundaries. At the outer boundary ($r_{max}$) we impose a so called zero-gradient
outflow condition. During periastron
when large spirals may extend beyond $r_{max}$ this condition will 
allow material to leave the system and not create numerical artifacts.
At the inner boundary we set a viscous outflow condition where the material may
flow through $r_{min}$ with the local (azimuthally averaged) viscous inflow referring
to an accretion disk in equilibrium.
No matter is allowed to flow back into the system and the mass of the disk will slowly
decline. To ensure a uniform setup for the planets we rescale the disk mass when inserting
them.

\begin{figure}[ht]
\def\capfrac{1}
\resizebox{0.47\linewidth}{!}{%
\includegraphics{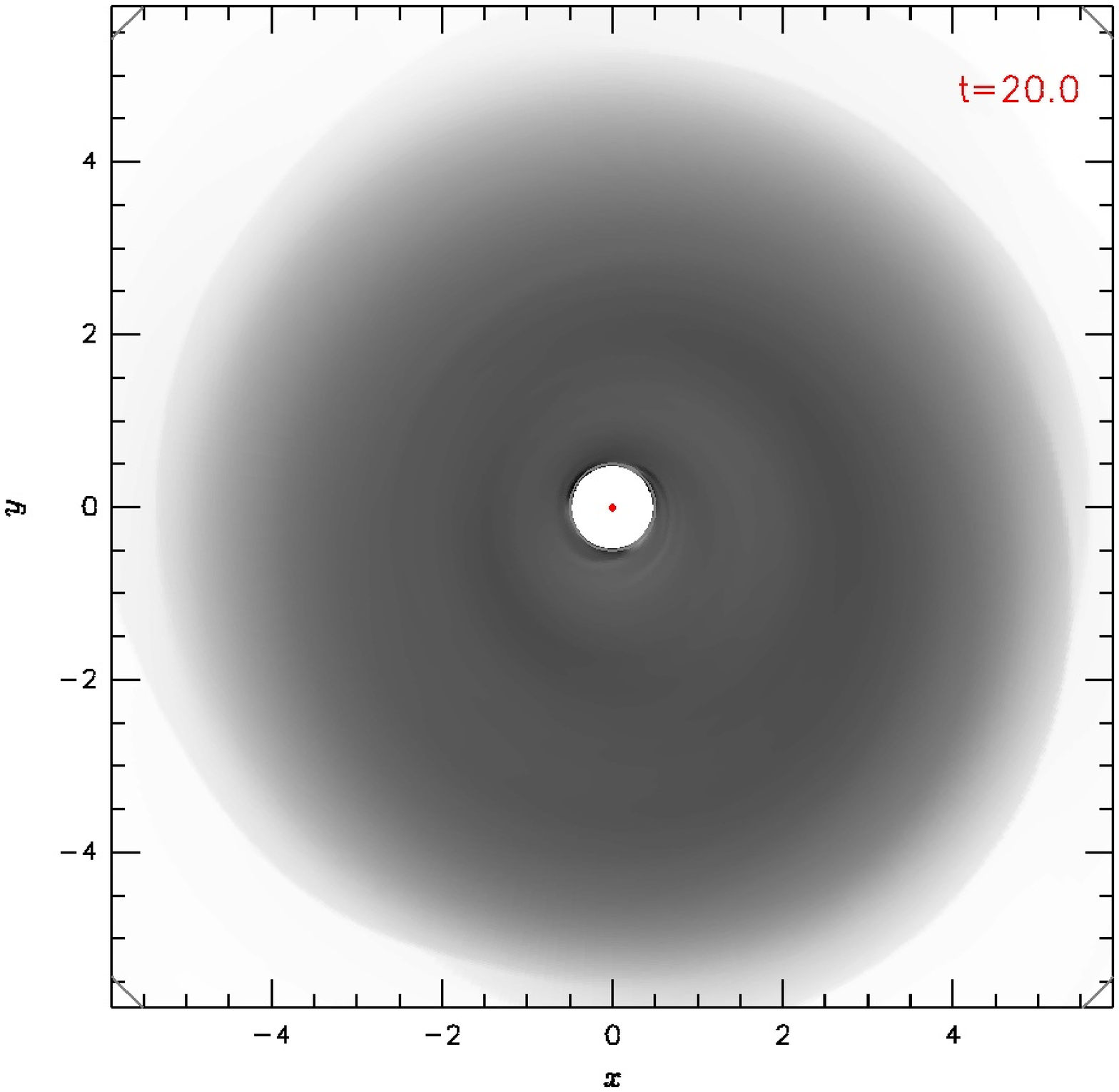}}
\resizebox{0.47\linewidth}{!}{%
\includegraphics{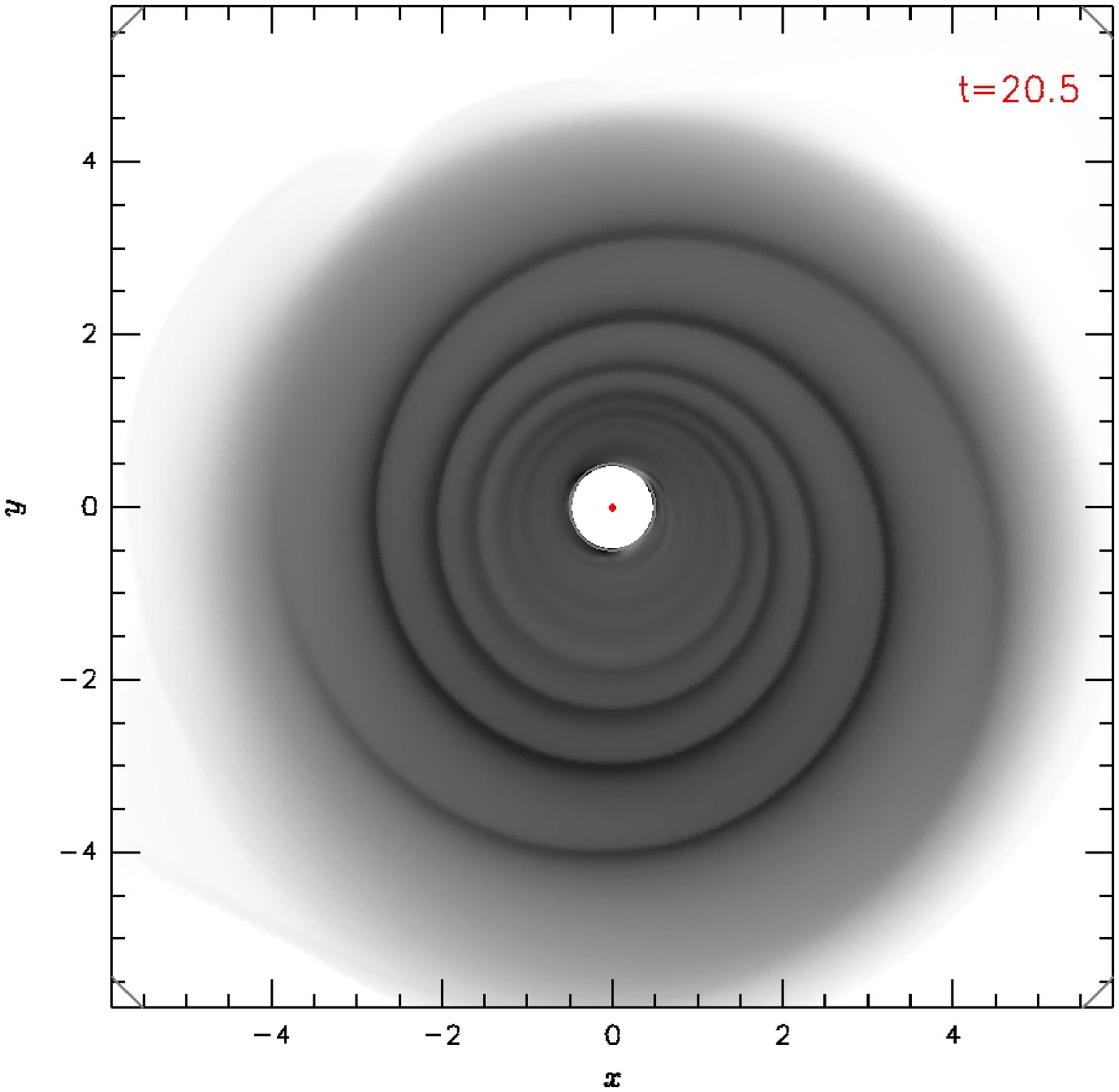}}
  \caption{
Grayscale plot of the two-dimensional density distribution
of the circumstellar disk around the primary at two different orbital
phases of the binary. 
{\bf Left} shortly after apocente at about 20 binary orbits, and
{\bf Right} shortly after closest approach (pericentre).
  }
   \label{wk-rpn-fig:h03-xy}
\end{figure}

\begin{figure}[ht]
\def\capfrac{1}
\resizebox{0.47\linewidth}{!}{%
\includegraphics{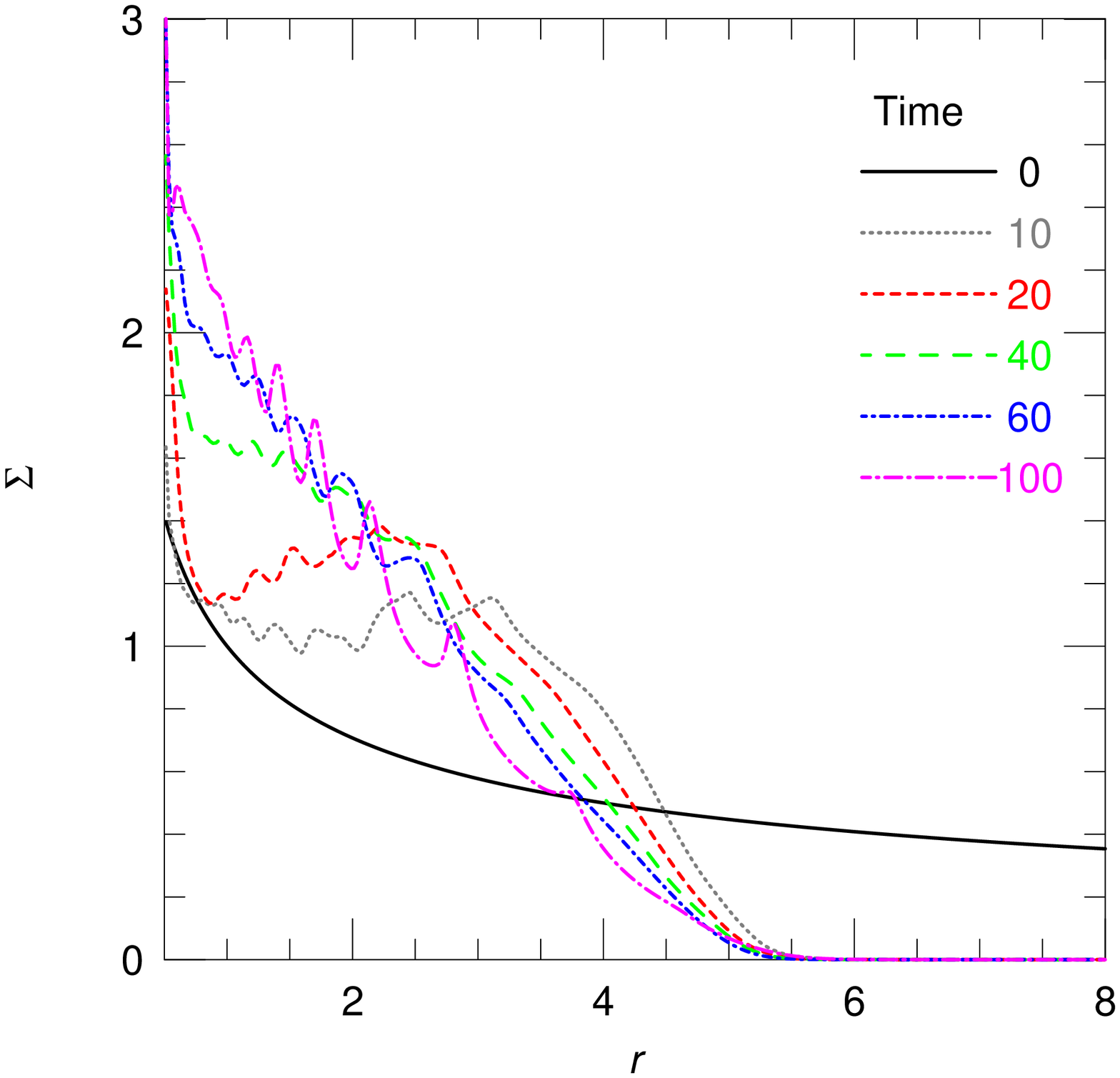}}
\resizebox{0.47\linewidth}{!}{%
\includegraphics{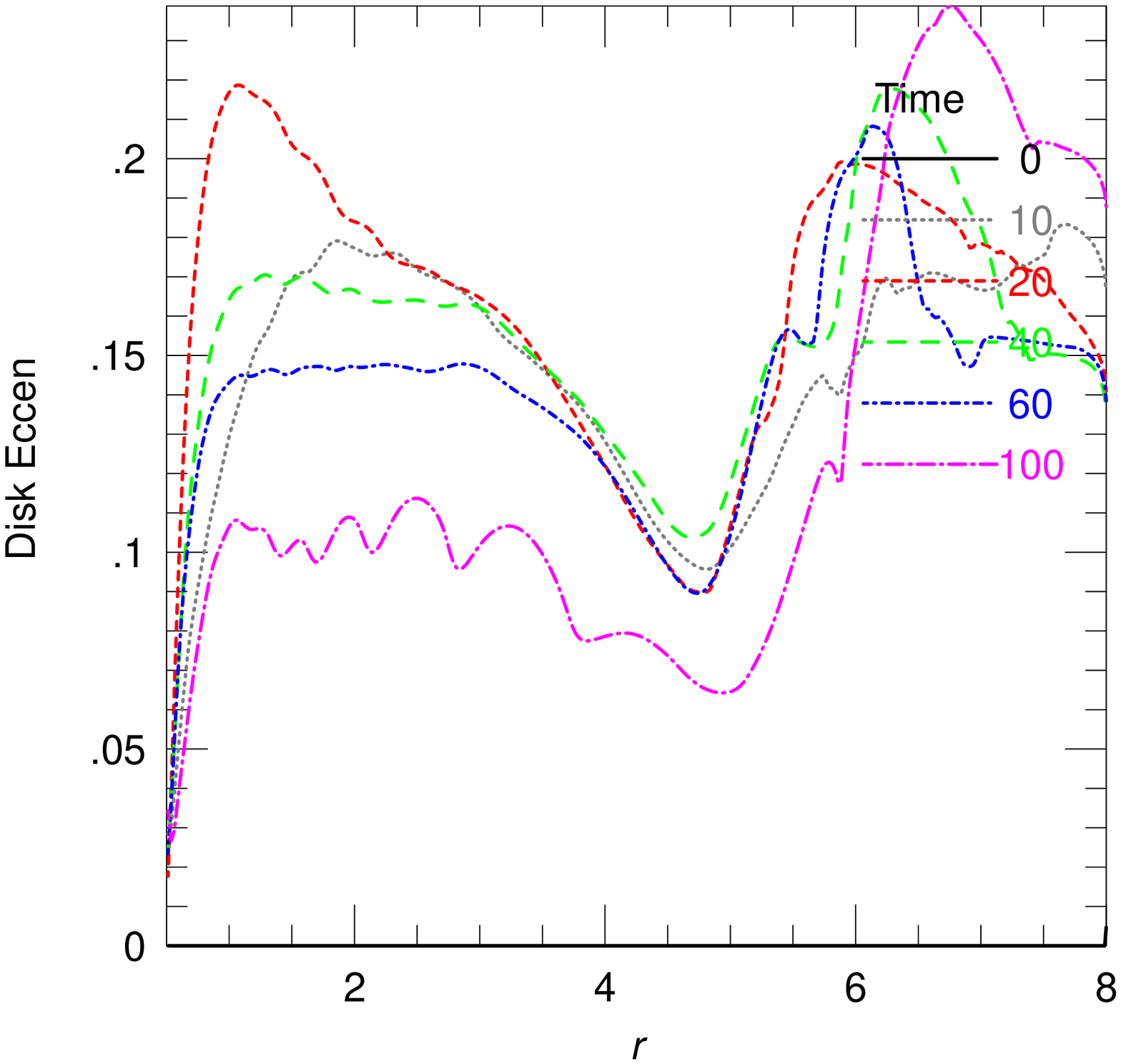}}
  \caption{
The radial surface density distribution ({\bf Left}) and the eccentricity 
({\bf Right}) of the circumstellar disk around the primary in the presence of
the secondary. Time is given units of the binary orbit, radial distance 
in AU, and the density in dimensionless units.
  }
   \label{wk-rpn-fig:sig-ecc-h03}
\end{figure}

\begin{figure}[ht]
\def\capfrac{1}
\resizebox{0.47\linewidth}{!}{%
\includegraphics{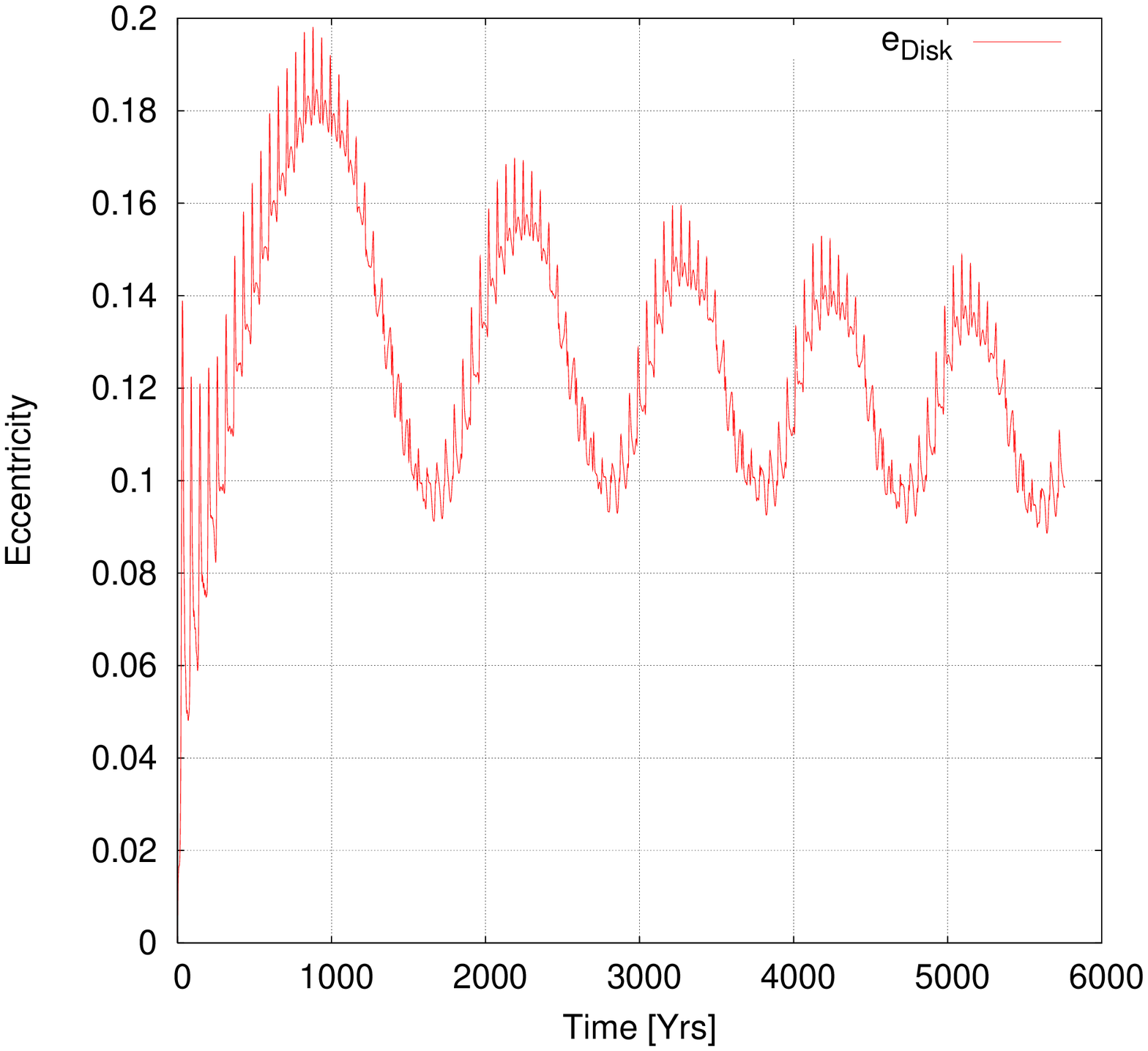}}
\resizebox{0.47\linewidth}{!}{%
\includegraphics{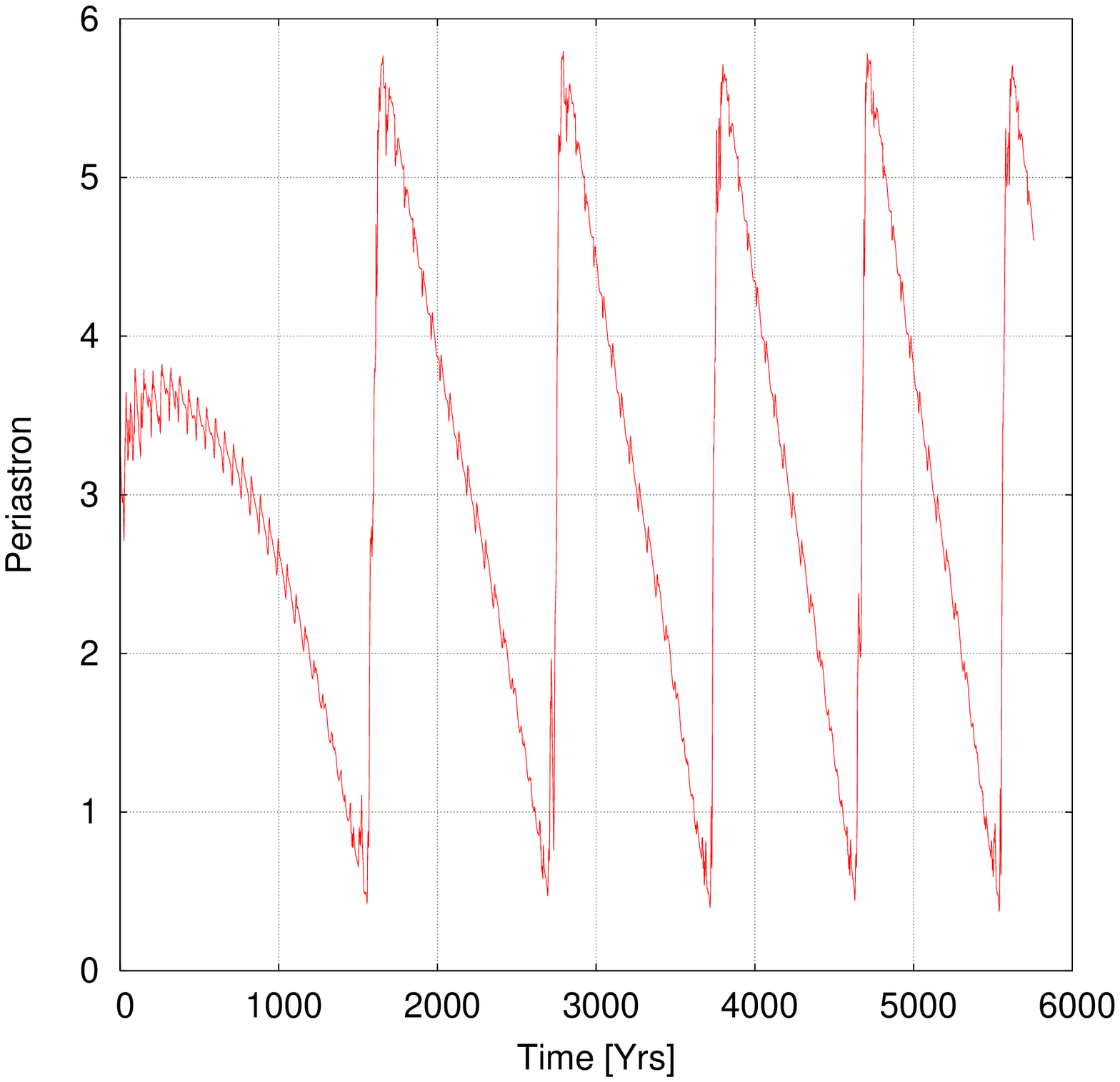}}
  \caption{
The evolution of the global mass averaged disk eccentriticy ({\bf left}) and
the position angle of the disk's periapse ({\bf right}).
  }
   \label{wk-rpn-fig:ecc-om-h03}
\end{figure}

\subsubsection{The structure of the disk} 
The presence of an eccentric secondary star leads to a strong periodic
disturbance of the disk whenever it is at periastron. Two strong
spiral arms (shock waves) are created in the disk which carry material
beyond the outer boundary of the computational domain. In between the
periapses the disk settles down and becomes more circular again.
This effect is illustrated in the Fig.~\ref{wk-rpn-fig:h03-xy}
where we display the surface density $\Sigma$ of the disk in gray scale
at 2 different times in the early evolution of the disk, see also
\citet{2000ApJ...537L..65N}.
Already the very first approaches with the binary lead to a truncation of
the disk as visible in left panel of Fig.~\ref{wk-rpn-fig:sig-ecc-h03}
for the curve at $t=10$ binary orbits. Slowly the whole disk structure
rearranges and equilibrates at around $t=50$ where it shows a much steeper
density slope than in the intial state.
The timescale for this equilibration process depends on the magnitude
of the disk viscosity.
The eccentricity of the disk in the final state of the disk varies approximately
between 0.1 and 0.16 depending on the position of the binary in its orbit
as shown in the left panel of Fig.~\ref{wk-rpn-fig:ecc-om-h03}.
The disk eccentricity $e_{disk}(r)$ has been obtained by calculating the 
eccentricity of each disk element, as if in a two body motion with the
primary star, and then averaged over the respective annulus.
At the same time the disk as a whole precesses as is shown in the right
panel of Fig.~\ref{wk-rpn-fig:ecc-om-h03}. This coherent slow retrograde
precession with a pattern speed much smaller than the orbital period of
the disk material around the star is caused by the non-negligible 
pressure forces operating in the disk. Similar behaviour has been demonstrated
for disks with free eccentricity \citep{2005A&A...432..757P}. 

\subsubsection{The orbital elements of the binary} 
In the previous section we have seen that the gravitational
torques of the binary lead to a truncation of the disk 
and re-arrangement of the material within. In turn, we expect a
change in the orbital elements of the binary.

To estimate theoretically the magnitude of the back reaction 
a circumstellar disk has on the orbital elements of the binary  
during the initial phase of readjustment,
we assume an idealized system consisting of a binary system
and a ringlike mass distribution
orbiting star 1 with mass $m_{ring}$, at a distance ($\delta$-function)
of $r_{ring}$. The energy $E_{bin}$ and angular momentum $L_{bin}$
of the binary is given by
\begin{equation}
\label{wk-rpn-eqn:el-bin}
   E_{bin} = - \, \frac{G M \mu}{2 a_{bin}}, \quad
   L_{bin} = \mu \left( G M a_{bin} \, ( 1 - e_{bin}^2) \right)^{1/2},
\end{equation}
and the corresponding quantities of the ring are 
\begin{equation}
   E_{ring} = - \, \frac{G M_1 m_{disk}}{2 r_{ring}}, \quad
   L_{ring} = m_{ring} \, \left( G M_1  r_{ring} \right)^{1/2},
\end{equation}
where $M = M_1 + M_2$ is the total mass of the two stars and 
$\mu = M_1 M_2 / M$ is the reduced mass.
Now, suppose that the ring is shifted from its initial position 
$r_{ring}^\alpha$ to a smaller radius $r_{ring}^\beta$ keeping all
its mass. This radius change mimicks the initial truncation of disk by the
binary. Through this process the ring's energy and angular momentum are
reduced from $E_{ring}^\alpha$ and $L_{ring}^\alpha$ to
$E_{ring}^\beta$ and $L_{ring}^\beta$.
By conservation of total enery and angular momentum
\begin{equation}
   E = E_{ring}  + E_{bin} \quad
   L = L_{ring}  + L_{bin},
\end{equation}
we can calculate the corresponding change in the orbital elements
of the binary from $E_{bin}^\alpha$ and $L_{bin}^\alpha$ to
$E_{bin}^\beta$ and $L_{bin}^\beta$.
For the binary paramter masses $M_1 = 1.6 M_\odot, M_2 = 0.4 \odot$ with initial
orbital elements $a_{bin}^\alpha =18.5$AU and $e_{bin}^\alpha=0.36$
we find for the shift of a ring with
 $m_{ring}=4 \cdot 10^{-3} M_\odot$  and initial radius $r_{ring}^\alpha = 4.0$AU
to a final radius of $r_{ring}^\beta = 2.0$AU that the binary elements change
to $a_{bin}^\beta =19.4$AU and $e_{bin}^\beta=0.41$.
A quite substantial change considering the smallness of the ring's mass
in comparision to the stellar masses. But the closeness to the primary
allows to gain a substantial amount of binding energy from the ring.
The calculation is approximate in the sense that the energy
and angular momentum of the ring are calculated with respect to
the non-inertial coordinate frame centered on the primary.

\begin{figure}[ht]
\def\capfrac{1}
\resizebox{0.99\linewidth}{!}{%
\includegraphics{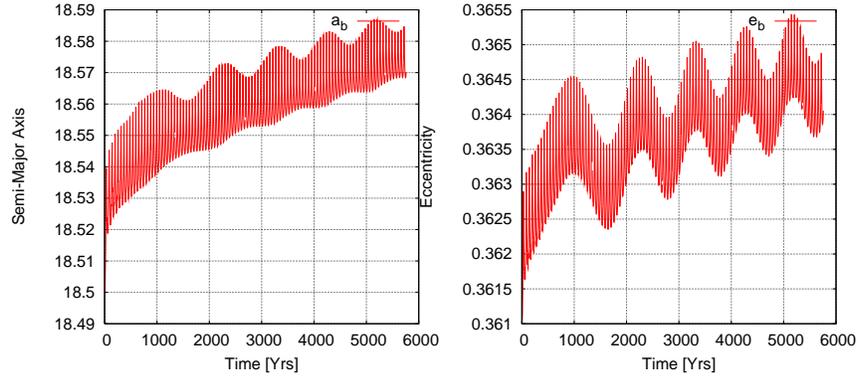}}
  \caption{
The evolution of the binary elements due to the interaction with the circumstellar
disk around the primary star, without an embedded planet.
One binary orbit refers to approximately 57yrs.
{\bf Left}: $a_{bin}(t)$; {\bf Right}: $e_{bin}(t)$.
  }
   \label{wk-rpn-fig:aeb-h03}
\end{figure}
We can now compare this estimate with the previous hydrodynamical simulations
and plot in Fig.~\ref{wk-rpn-fig:aeb-h03} the evolution of $a_{bin}$ and $e_{bin}$
for about the first 100 binary periods with no planet included.
As demonstrated above, the binary expands as it gains energy from the compressed disk
and increases its eccentricity. The increase in $e_{bin}$ does not lead to
a decrease in the angular momentum however, since it increases its separation, see
Eq.~\ref{wk-rpn-eqn:el-bin}.
Whenever the binary is near periastron the gravitational interaction with the
disk is maximal which results in the strong periodic spikes in the binary elements.
The change in the orbital elements of the binary is somewhat smaller than
the estimated values because {\it i}) the mass of disk is smaller in the hydrodynamic
calculation and {\it ii}) disk mass and angular momentum are stripped off
by the secondary and are lost through the outer boundary of the computational domain.
The loss through the (open) inner boundary of the disk is only marginal.

\subsubsection{The behaviour of an embedded planet} 
In the previous section we have seen that the gravitational
torques of the binary lead to a truncation of the disk and a rearrangement
of the disk material. To study the influence of the companion on 
the evolution of small protoplanets we embed, after an equilibration
time of 100 binary orbits (nearly 6000 yrs), a $30 M_{Earth}$ planet in
the disk and follow its subsequent evolution. 
This rather time consuming procedure to generate the initial
state is necessary to obtain realistic initial conditions for the
growing protoplanet.
At the time of insertion of the planet the remaining disk mass is rescaled
to contain 3 $M_{Jup}$ within the computational domain.

\begin{figure}[ht]
\def\capfrac{1} 
\resizebox{0.47\linewidth}{!}{%
\includegraphics{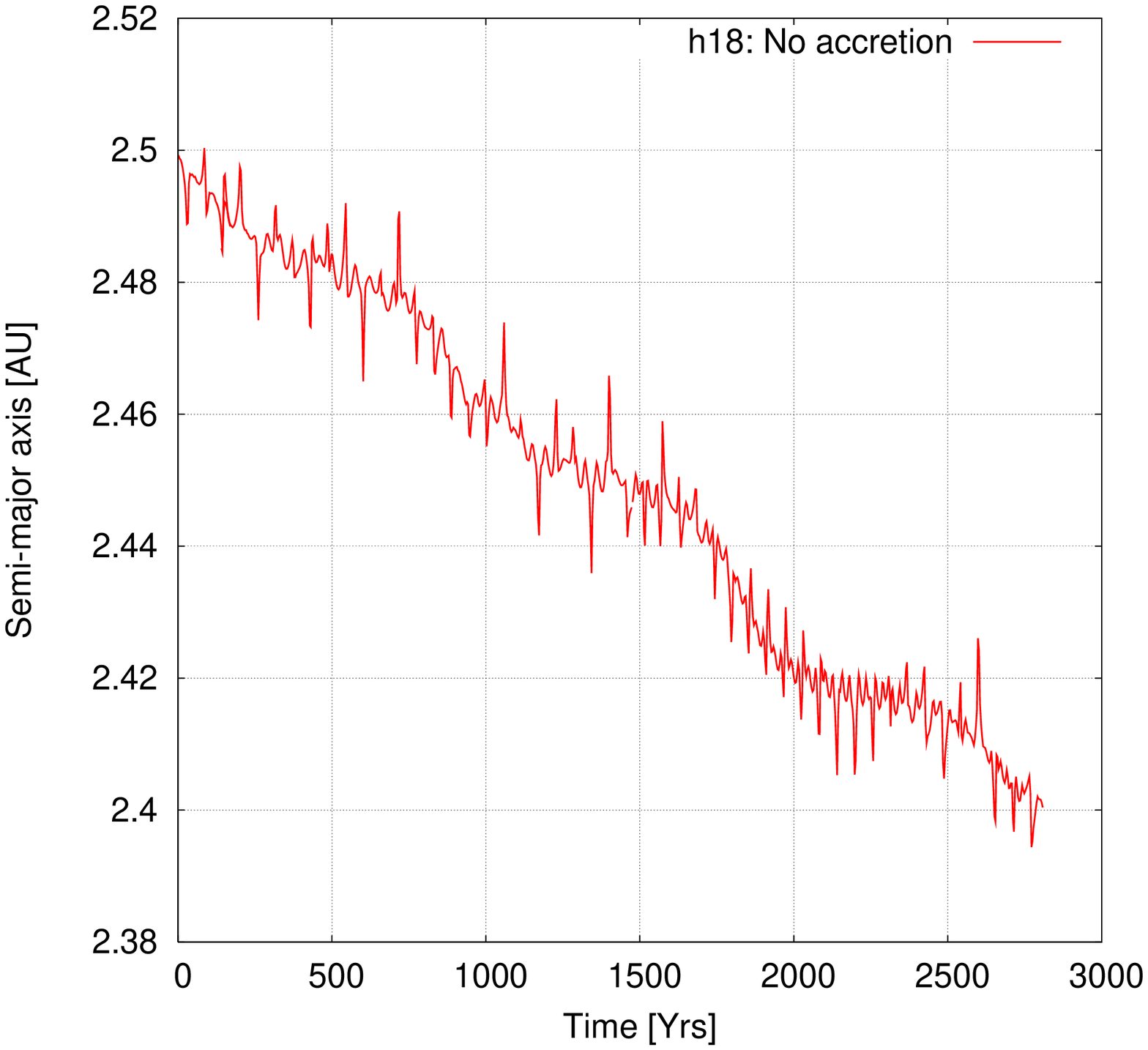}} 
\resizebox{0.47\linewidth}{!}{%
\includegraphics{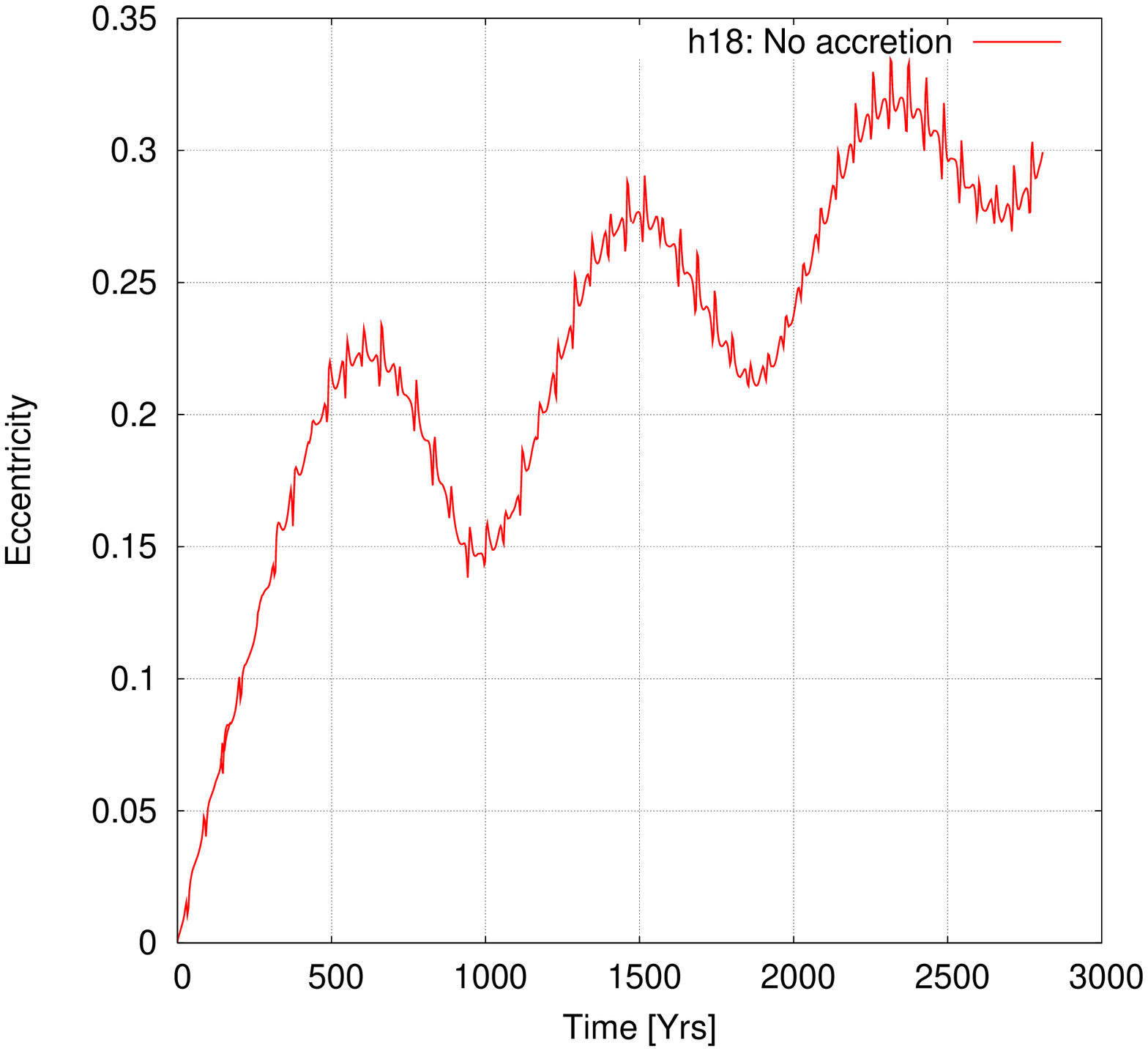}}
  \caption{
The evolution of the semi-major ({\bf left}) axis and eccentricity 
({\bf right}) of an embedded planet in the circumstellar accretion disk.
Here, the planet is not allowed to accrete material from the disk
and remains at 30 $M_{Earth}$. The planet is inserted after 100 
orbital binary periods, and the time is reset to zero.
  }
   \label{wk-rpn-fig:aep1-h18}
\end{figure}

As a first sample case we follow the planet's orbital evolution while
keeping its mass constant, i.e. the planet is not allowed to accrete
mass from its environment.
This model will serve as a reference for the subsequent cases which
will allow for planetary mass growth.
The planet is released at $a_p = 2.5$AU on a circular orbit.
After insertion of the planet its orbital elements will change due to 
gravitational interaction with the disk and the binary.
The planet migrates inward due to the torques of the disk,
with a rate of 0.1 AU in about 2800 yrs. While the overall migration
is approximately linear over this time, it is modulated by the binary
companion and the precessing, eccentric disk (see left
panel of Fig.~\ref{wk-rpn-fig:aep1-h18}).
At the same time the planetary  eccentricity increases to about 0.3, 
with the eccentric disk yielding the prime contribution to the growth
of $e_p$. The oscillatory behaviour originates from the changing degree
of apsidal alignent between eccentric disk and planet as they undergo
relative precession.

\begin{figure}[ht]
\def\capfrac{1}
\resizebox{0.70\linewidth}{!}{%
\includegraphics{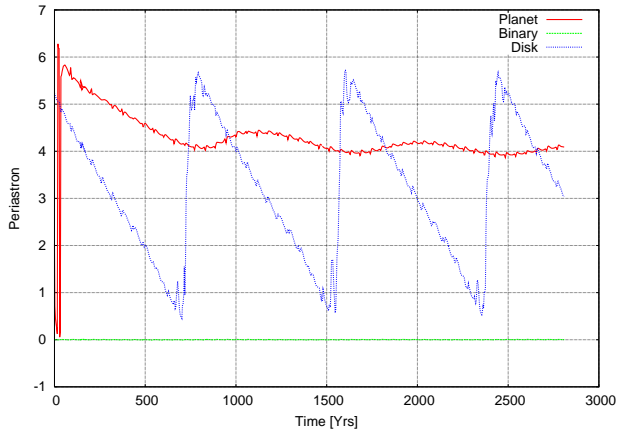}}
  \caption{
The evolution of the argument of pericenter of the disk, the planet and the 
binary after insertion of a 30 $M_{Earth}$ planet.
  }
   \label{wk-rpn-fig:omdp-h18}
\end{figure}
The evolution of the argument of pericenter of
the disk, the planet and the binary are displayed 
in Fig.~\ref{wk-rpn-fig:omdp-h18}. While the disk continues its
retrograde precession and the binary remains unchanged, the planet undergoes
initially a retrograde precession and then settles to an approximately
constant value with oscillations whose frequency is given by the
precession frequency of the whole disk in which it is embedded.

\begin{figure}[ht]
\def\capfrac{1}
\resizebox{0.70\linewidth}{!}{%
\includegraphics{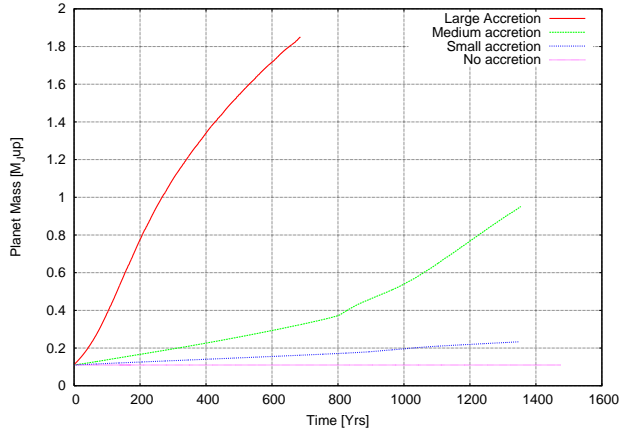}}
  \caption{
The evolution of the argument of pericenter of the disk, the planet and the 
binary after insertion of a 30 $M_{Earth}$ planet.
The planets are inserted after 100
orbital binary periods, and the time is reset to zero.
  }
   \label{wk-rpn-fig:mp-h16-h18}
\end{figure}

\begin{figure}[ht]
\def\capfrac{1}
\resizebox{0.47\linewidth}{!}{%
\includegraphics{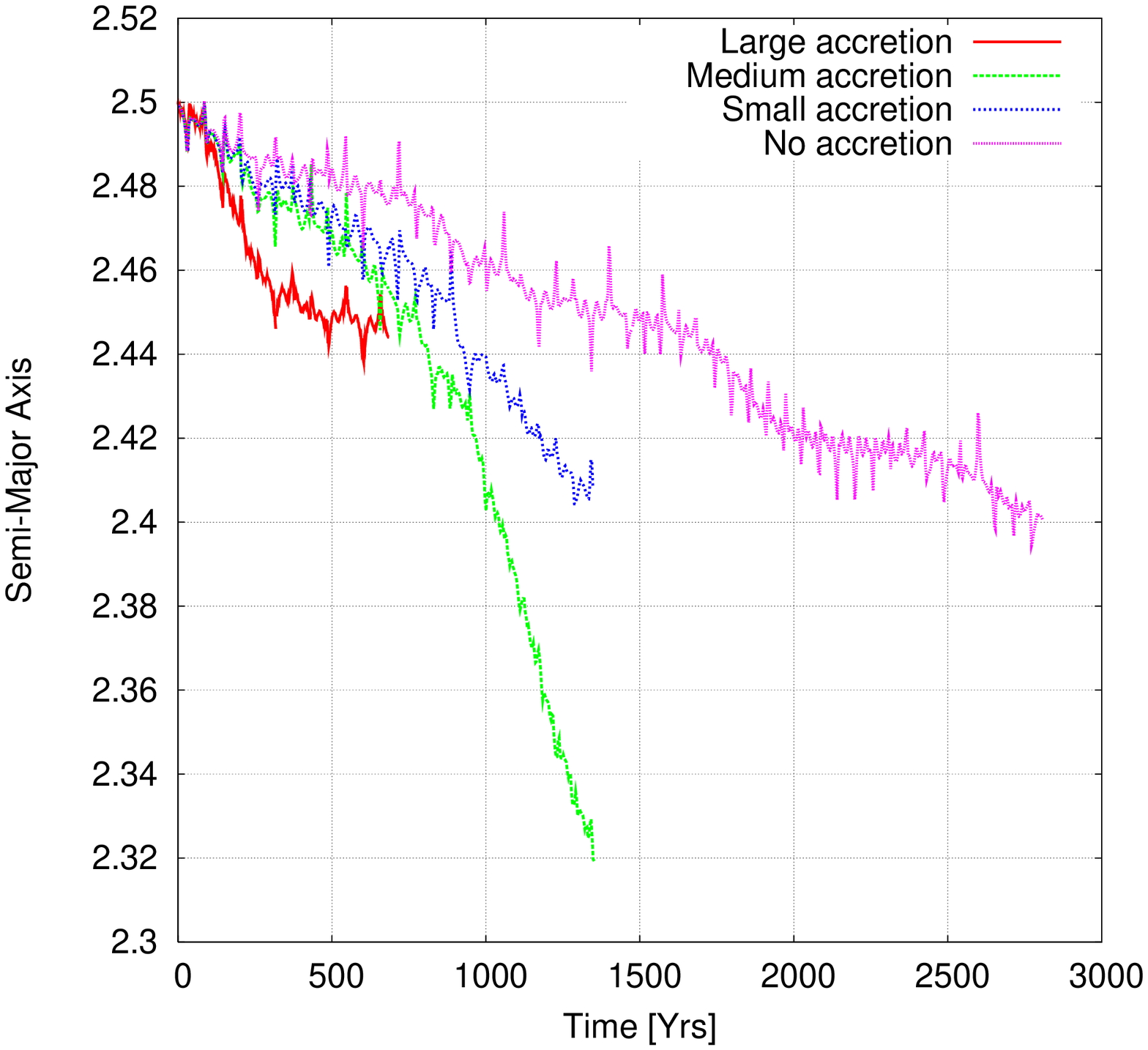}} 
\resizebox{0.47\linewidth}{!}{%
\includegraphics{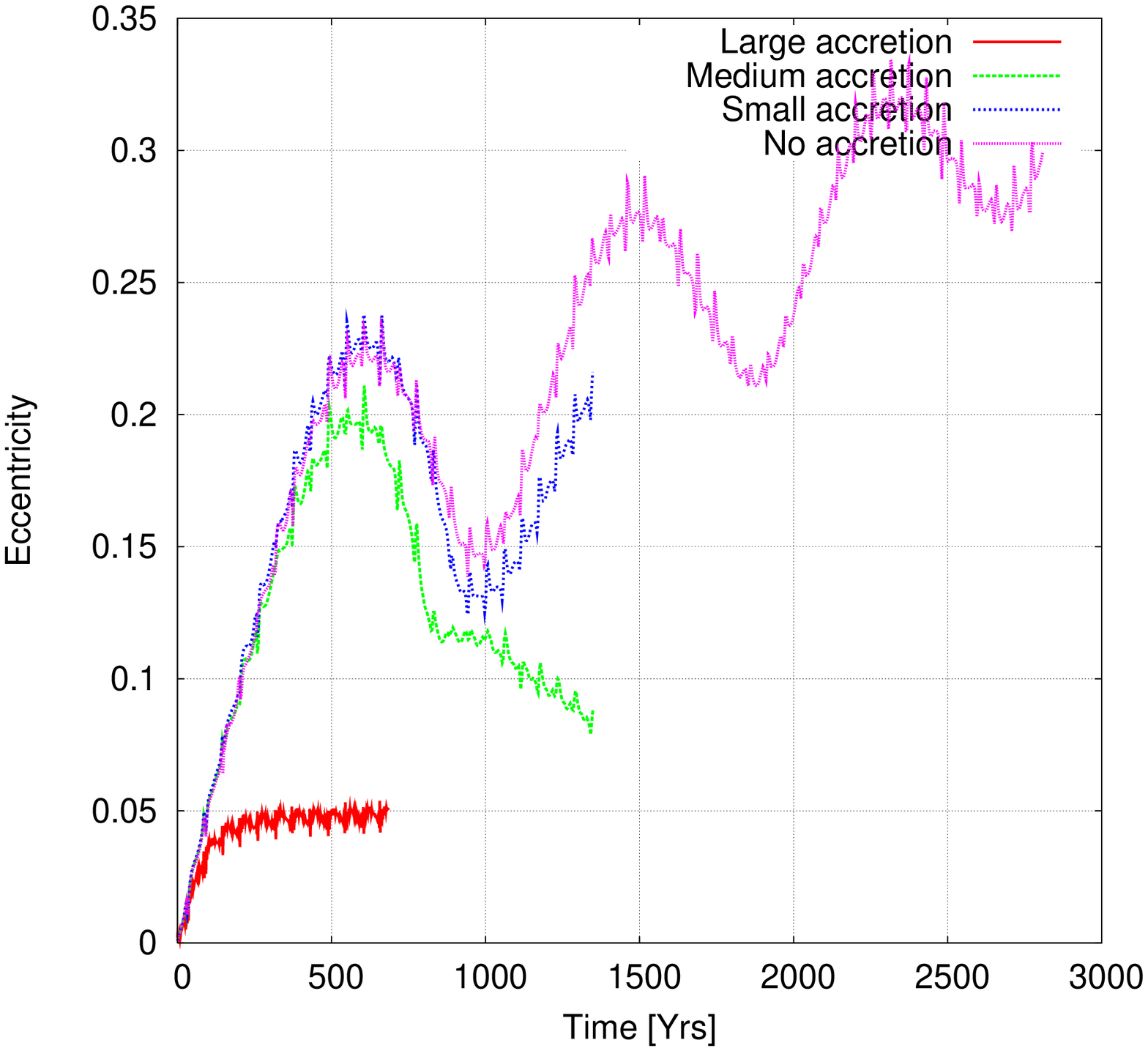}}
  \caption{
The evolution of the semi-major ({\bf left}) axis and eccentricity
({\bf right}) of embedded planets in the circumstellar accretion disk.
The planets all start at the same mass but accrete at different rates from
the accretion disk.
The planets are inserted after 100
orbital binary periods, and the time is reset to zero.
  }
   \label{wk-rpn-fig:aep-h16-18}
\end{figure}

To study more realistic cases we now allow the planet to grow in mass from
the disk during its motion through it.
The accretion process is modelled numerically in a simple manner. At each time
step a certain fraction of the material within the Roche lobe of the planet is
taken out of the computational domain and added to planet's mass.
In Fig.~\ref{wk-rpn-fig:mp-h16-h18} we show the evolution of the mass
of the planet for different accretion rates.
For the largest accretion rates the planet aquires over $1.8 M_{Jup}$ within
the first 700 yrs of its evolution, a value that is unrealistically high.  
So this model sets the limiting case for the others.
The model with the small accretion only doubles its mass from 30 to
60 $M_{Earth}$ during the first 1000 yrs which gives a more realistic
accretion rate.
The no accreting case is given by the horizontal line.

More interesting is now the different orbital behaviour of the planets
which is displayed in Fig.\ref{wk-rpn-fig:aep-h16-18}.
The planet with the constant mass has the slowest migration, and the larger
the accretion rate the larger is the migration speed.
This is consistent with the estimated migration rates for different masses
\citep{2003ApJ...586..540D}. The planet with the maximum accretion rate grows
rapidly in mass and approaches already after 280 yrs the 1 $M_{Jup}$
limit, when its migration rate slows down and levels off as the mass
in the disk decreases and the driving agent disappears.
The intermediate cases migrate initially with the same speed as the
non-accreting model but accelarate as the planetary mass increases.

Concerning the eccentricity evolution, the lightest 
planet experiences the largest
growth. For the large accretion rate the eccentricity soon levels off
to a value of $e_p =0.05$.

\subsubsection{Comparison with $\gamma$ Cep}
The most up to date observational data suggest the following
parameters for the planet in the $\gamma$ Cep system:
$a_p \simeq 2.044$, $e_p \simeq 0.115$ and 
$m_p \sin{i} \simeq 1.60$ M$_{Jupiter}$. If this planet formed 
according to the core instability model, then our simulations
raise a number of important questions that we are currently addressing.

First, a low mass, non accreting planet embedded in an 
the eccentric disk experienced substantial growth in eccentricity
(see Fig.~\ref{wk-rpn-fig:aep1-h18}).
This has clear implications for the accretion of planetesimals
because their velocity dispersion may become very large due to this
effect. \citet{2004A&A...427.1097T} examined the evolution of
planetesimal orbits under the influence of the binary companion
and aerodynamical gas drag. They concluded that accretion of 
planetesimals would occur in the shear dominated regime
because orbital alignment was maintained due to the gas drag. 
This work, however, did not include the effects of an eccentric
disk, and so it remains unclear whether planetesimal orbits will
remain aligned. We will discuss the effects of including the full
dynamics of the disk when calculating the orbital evolution of planetesimals
in the $\gamma$ Cep system in the next section. 

A second issue is that of type I migration of the giant planet core
that must survive before gas accretion occurs. Fig.~\ref{wk-rpn-fig:aep1-h18}
shows the non  accreting, low mass planet undergoing quite rapid inward
migration. The migration, however, is modulated by the eccentricity
of the planet, such that at high eccentricity phases the migration rate
decreases. It is possible that longer run times will show an
essential stalling of this migration if the planet eccentricity
grows beyond its final value of $e_p \simeq 0.3$. Simulations are
currently being conducted to examine this in more detail.

Once gas accretion is switched on, it is clear that a disk mass of
about 3 Jupiter masses, where the outer disk radius is tidally
truncated at $r \simeq 5$ AU, will be sufficient to grow a planet
that is close to the minimum observed mass of $m_p \sin{i} \simeq 2.044$ 
M$_{Jupiter}$. It is also clear that we can construct a model
in which a low mass planet growing
from an initially circular orbit can achieve a final mass of $m_p \simeq 2$
M$_{Jupiter}$, and have a final eccentricity of $e_p \simeq 0.1$ as required.
Calculations are underway to see if a planetary core on an initially
eccentric orbit (as expected from Fig.~\ref{wk-rpn-fig:aep1-h18}),
will circularise as it accretes gas from the disk such that a 
self consistent model that fits the observations can be constructed.

A final comment relates to the final mass of the planet.
Our simulations suggest that a disk mass of about 3 Jupiter masses
will be enough to form a gas giant of the required miminum mass.
A future test of the mode by which the planet in $\gamma$ Cep formed
(gravitational instability versus core accretion)
will be determination of its actual mass. We suspect that a disk
that is massive enough to form a planet through gravitational
unstability will lead to a planet whose final mass is
substantially larger than the minimum value observed.

\section{Evolution of planetesimals in a circumstellar disk
with a companion}
\begin{figure}[ht]
\def\capfrac{1}
\resizebox{0.99\linewidth}{!}{%
\includegraphics{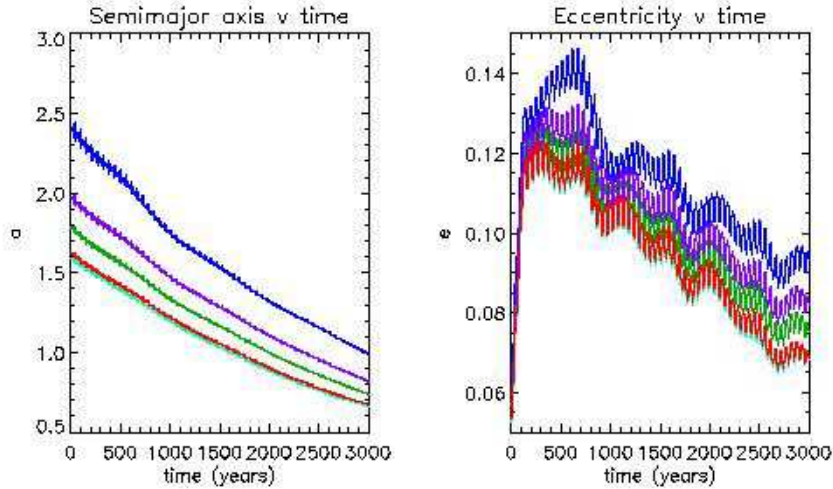}}
  \caption{
The evolution of the semi-major axes (left panel) and eccentricities
(right panel) of embedded planetesimals in the circumstellar accretion disk.
  }
   \label{aeplanetesimals}
\end{figure}
We now describe preliminary results from simulations of planetesimals
embedded in circumstellar disks with a companion star. We take as
our basic model the disk and binary system described in the
previous section~\ref{sec:circumstellar}. As in the models in which
low mass protoplanets were considered, we evolve the system for
100 binary orbits prior to inserting 100 planetesimals. 
At the point when the planetesimals are inserted, the disk mass is augmented
so that it contains 3 Jupiter masses in total. The planetesimals are randomly
distributed initially between orbital radii of 1.5 and 2.5 AU 
on circular Keplerian orbits. We consider here planetesimals 
whose physical radii are 100 metres. A broader range of sizes will be 
discussed in Nelson \& Kley (2007, in preparation).
The planetesimals experience aerodynamic gas drag using
the standard formulae found in \citet{1977MNRAS.180...57W},
and also experience the gravitational
force due to the disk, central star and companion star. 
Although the simulations we describe here
are two dimensional, we assume that the planetesimals lie in
the disk midplane and calculate the volumetric density from
the surface density by assuming that the vertical density profile
is Gaussian with scale height $H=0.05 r$, where $r$ is the orbital radius.
We use linear interpolation to calculate the gas density and
velocity at the planetesimal positions for use in the gas drag formula.
\begin{figure}[ht]
\def\capfrac{1}
\resizebox{0.99\linewidth}{!}{%
\includegraphics{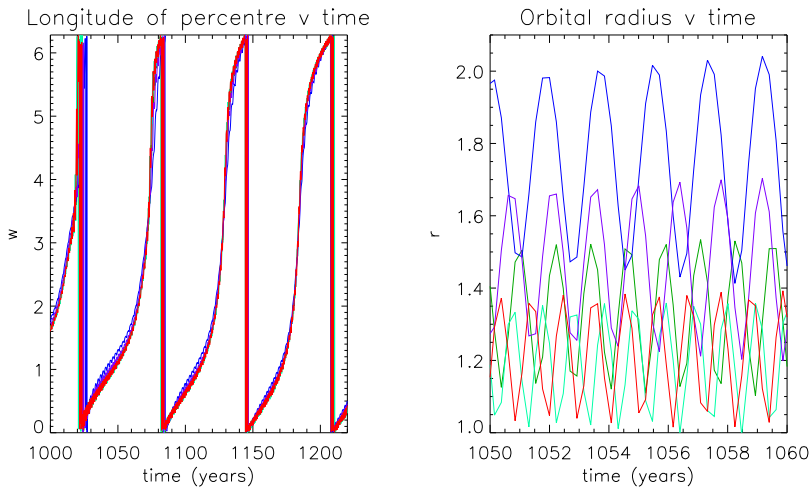}}
  \caption{
The evolution of the longitudes of pericentre (left panel) and orbital radii
(right panel) of embedded planetesimals in the circumstellar accretion disk.
Notice that the orbits cross one another, suggesting that high velocity
impacts are likely to occur.
  }
   \label{orbit-planetesimals}
\end{figure}
The evolution of the semi-major axes and eccentricities for
5 representative planetesimals are shown in figure~\ref{aeplanetesimals}.
We see that the planetsimals migrate inward on the expected
time scale due to the aerodynamic gas drag, and are also excited onto
orbits with high eccentricity ($e \ge 0.12$). The eccentricity is driven
upward primarily by gravitational interaction with the eccentric
gas disk, and not because of direct interaction with the binary companion.
As the planetesimals drift inward their eccentricity decays slightly
but still remains significant.

In the left panel of figure~\ref{orbit-planetesimals} we plot the 
longitude of pericentre of the five representative planetesimals
for times between 1000 and 1200 years after the planetesimals
were inserted.
We see that their orbits remain quite close to alignment, but
the alignment is not perfect and the degree of alignment is time
dependent. The right panel shows the orbital radii of the five
planetesimals, and we see clearly that the orbits cross.
Given eccentricities on the order of $e \simeq 0.1$ and semimajor
axes approximately $a \simeq 1.5$ AU, this suggests that collision
velocities between the planetesimals will be on the order
of 2 km s$^{-1}$. Simulations of colliding icy bodies with
radii $\simeq 100$ m performed by Benz \& Asphaug (1999) suggest
that disruption occurs for impact velocities $\simeq 15$ m $s^{-1}$,
a factor of $\simeq 1/133$ smaller than the velocity dispersions 
obtained in our simulations.
Clearly this raises questions about the applicability of the
core instability model when applied to close binary systems such
as $\gamma$ Cep, as it would appear that impacts between
planetesimals will be destructive rather than accretional.

\section{Evolution of planets in circumbinary disks} 
\label{circumbinary}
In this section we present the results of simulations
that examine the evolution of both low and high mass protoplanets 
which form in circumbinary disks. A fuller discussion of the work relating
to low mass planets is presented in \citet{2007MNRAS..submitted},
and a detailed description of the simulations relating
to high mass planets is presented in \citet{2003MNRAS.345..233N}.

We consider the interaction between a coplanar binary and protoplanet
system and a two--dimensional, gaseous, viscous, circumbinary disk
within which it is supposed the protoplanets form. We do not
address the formation process itself, but rather assume that
circumbinary protoplanets can form, and examine the dynamical 
consequences of this.
Each of the stellar components and the protoplanet experience
the gravitational force of the other two, as well as that due to the disk.
The planet and binary orbits are evolved using a fifth--order
Runge--Kutta scheme (Press et al. 1992). The force of the planet on
the disk, and of the disk on the planet, is softened
using a gravitational softening parameter $b=0.5a_p(H/r)$, where $a_p$ is the
semimajor axis of the planet, and $H/r$ is the disk aspect ratio.
We assume that the mass of the protoplanet is fixed,
and disk models have effective aspect ratio $H/r=0.05$.

\subsection{Low mass circumbinary planets}
The simulation described below was performed using the
hydrodynamics code {\tt GENESIS} \citep{2005sf2a.conf..733P, 2006MNRAS.370..529D}.
The Shakura--Sunyaev viscosity parameter $\alpha=2 \times 10^{-4}$, and
the disk was initialised to have a mass of 0.04 M$_{\odot}$ 
within a radius of 40 AU.
An expanded version of the following discussion is presented
in \citet{2007MNRAS..submitted}. 

\begin{figure}[ht]
\resizebox{0.47\linewidth}{!}{%
\includegraphics{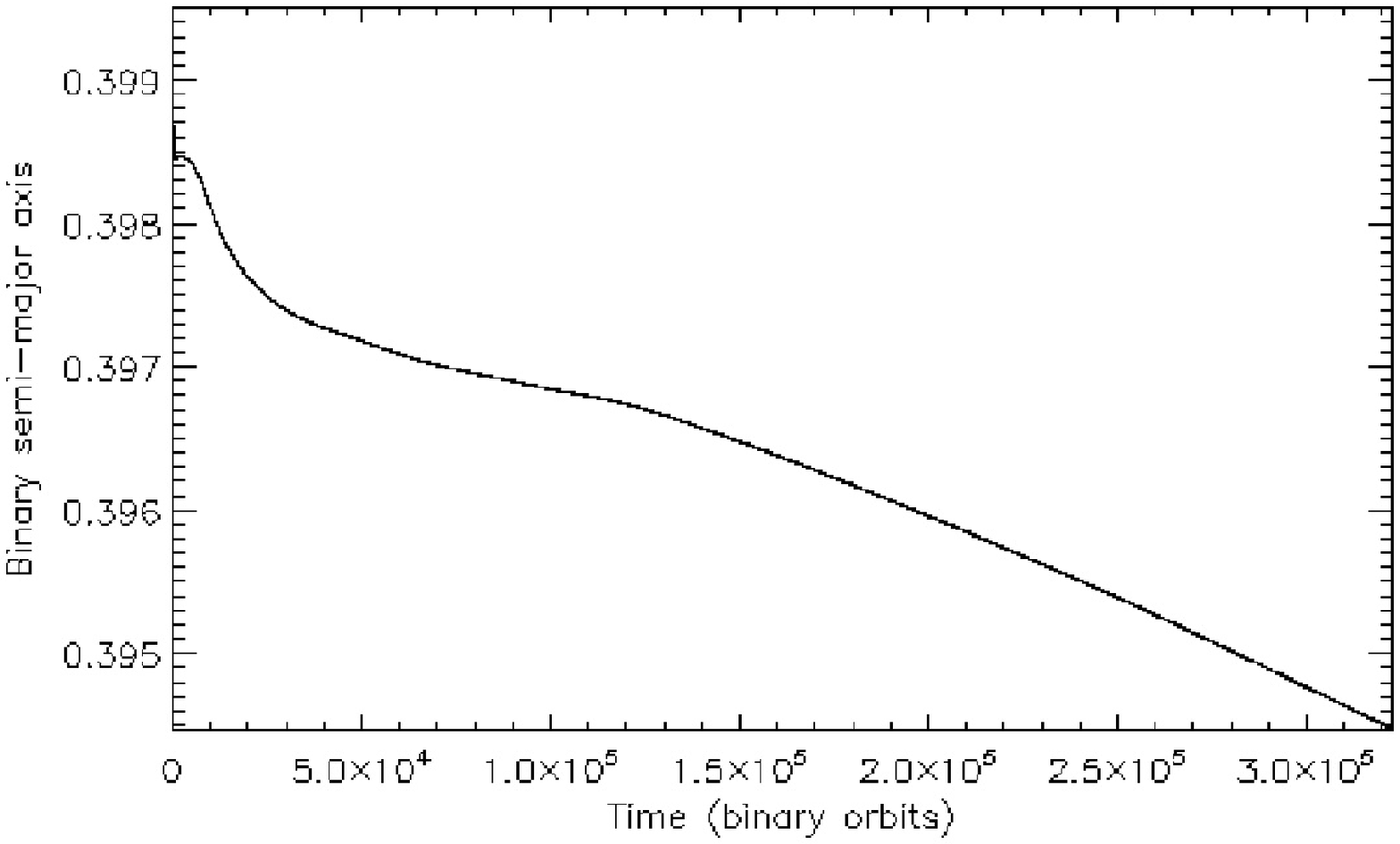} }
\resizebox{0.47\linewidth}{!}{%
\includegraphics{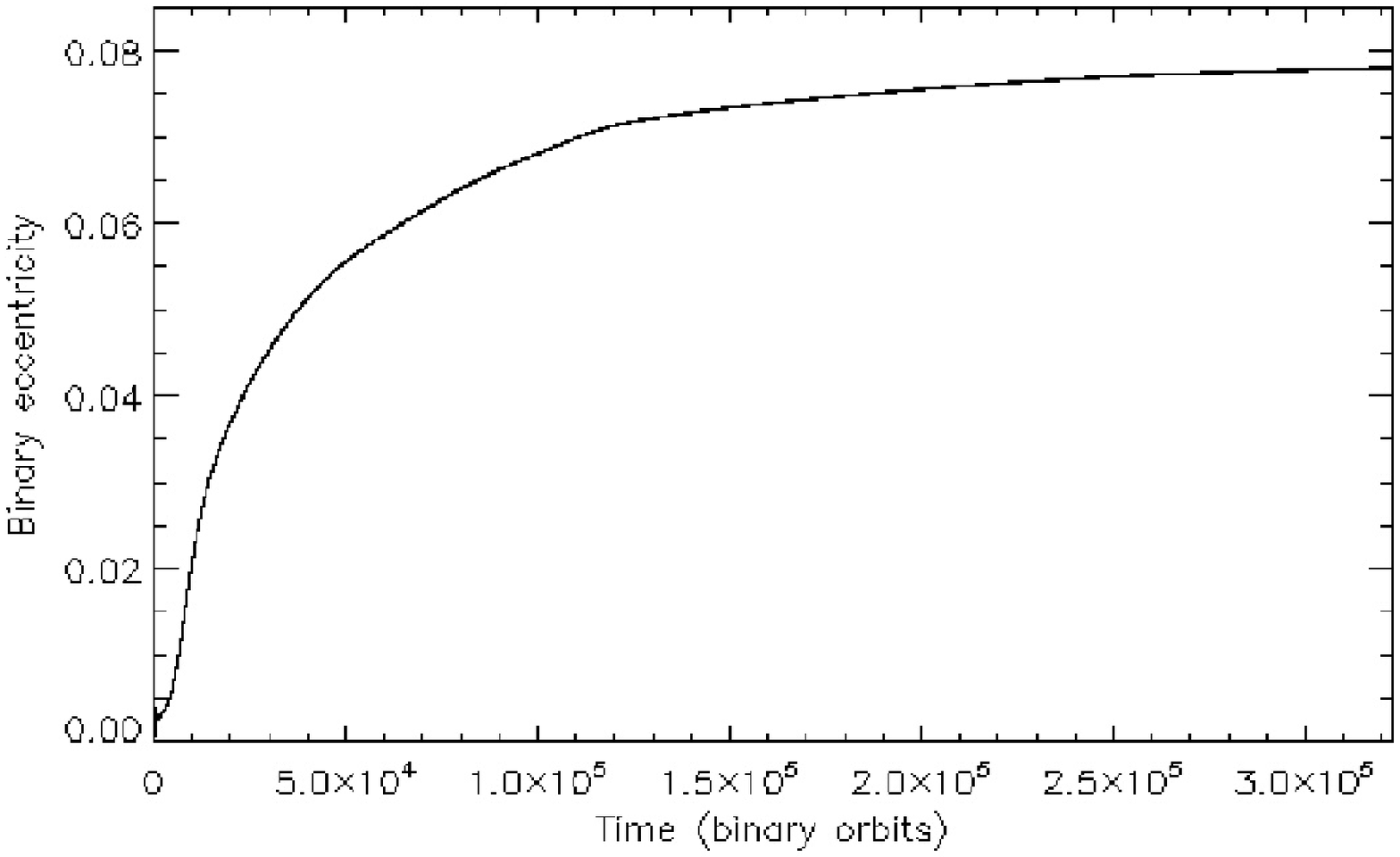} }
\caption[]
{The evolution of the binary elements due to interaction 
with the circumbinary disk. The left panel shows the semimajor
axis evolution over time (expressed in binary orbits),
and the right panel shows the eccentricity evolution. The binary
orbital period is $\sim 92$ days.}
\label{circbin1}
\end{figure}

The simulation was initialised with a binary star system on
a circular orbit surrounded by an unperturbed circumbinary
disk. The stellar masses were $M_1=1/11 M_\odot$
and $M_2=1/110 M_\odot$ (i.e. the mass ratio was $q=0.1$), and 
the semimajor axis $a_{bin}= 0.4$ AU.
The left panel of figure~\ref{circbin1} shows the slow decline
of the binary semimajor axis over a time scale of about 80,000 years
(the binary orbital period is approximately 92 days)
and the right panel shows the growth and saturation of the binary
eccentricity. As expected, interaction with the disk drives
the growth of binary eccentricity (e.g. Papaloizou, Nelson \& Masset 2001),
which eventually saturates at a value of $e_{bin} \simeq 0.08$.

\begin{figure}[ht]
\resizebox{0.47\linewidth}{!}{%
\includegraphics{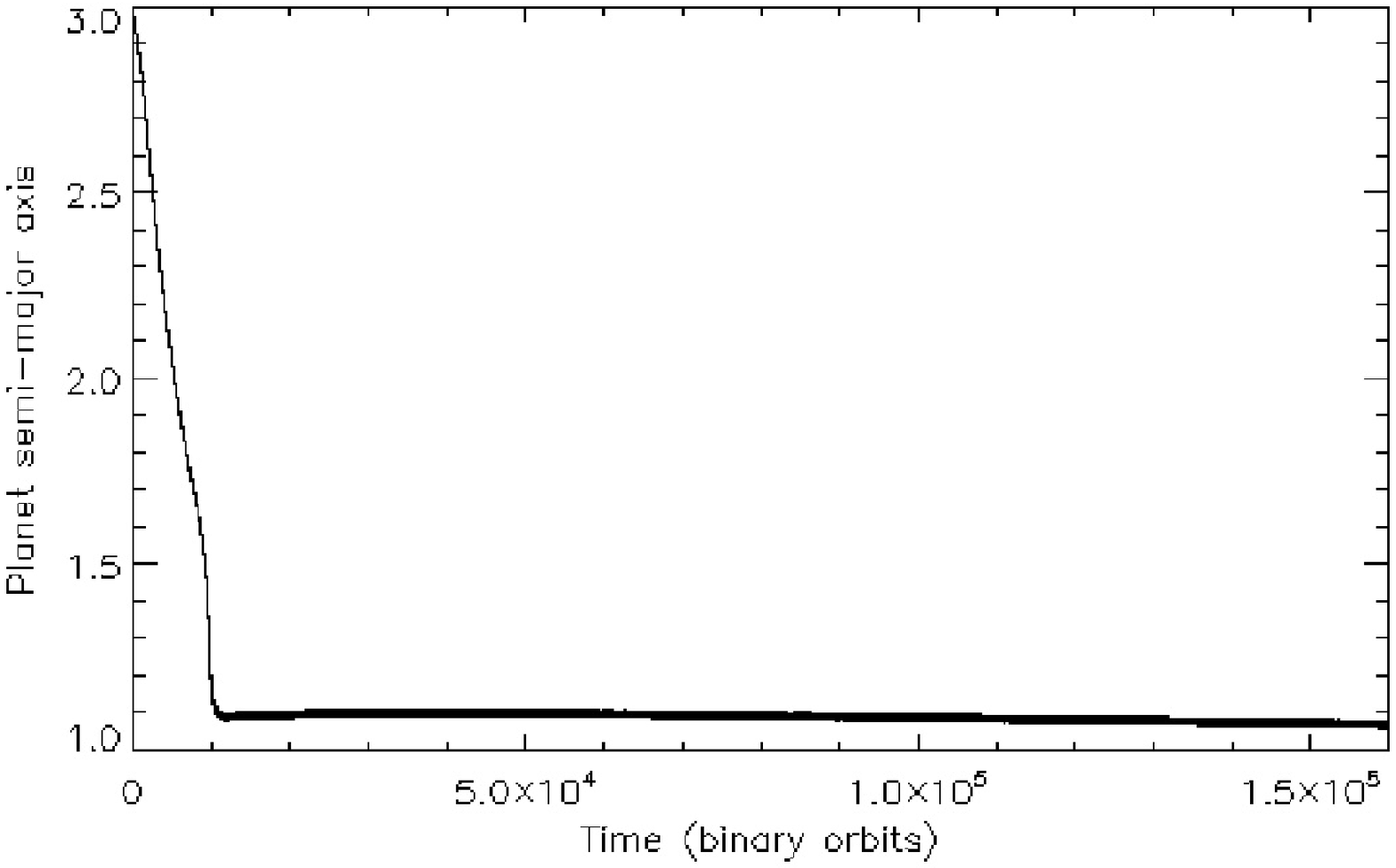} }
\resizebox{0.47\linewidth}{!}{%
\includegraphics{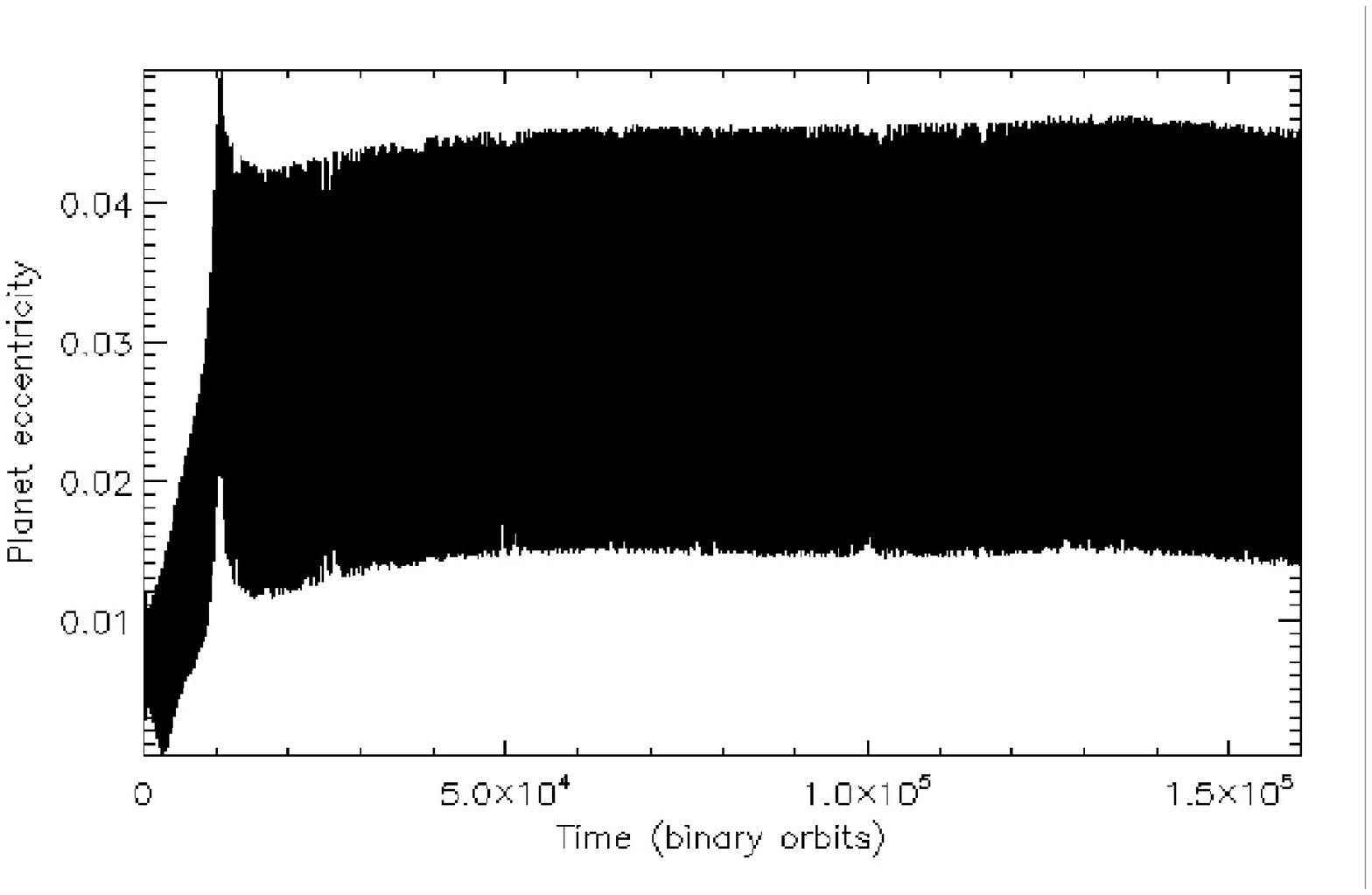} }
\caption[]
{The evolution of the planet elements due to interaction 
with the circumbinary disk. The left panel shows the semimajor
axis evolution over time in years, and the right panel shows the
eccentricity evolution.}
\label{circbin2}
\end{figure}

\begin{figure}[ht]
\resizebox{0.47\linewidth}{!}{%
\includegraphics{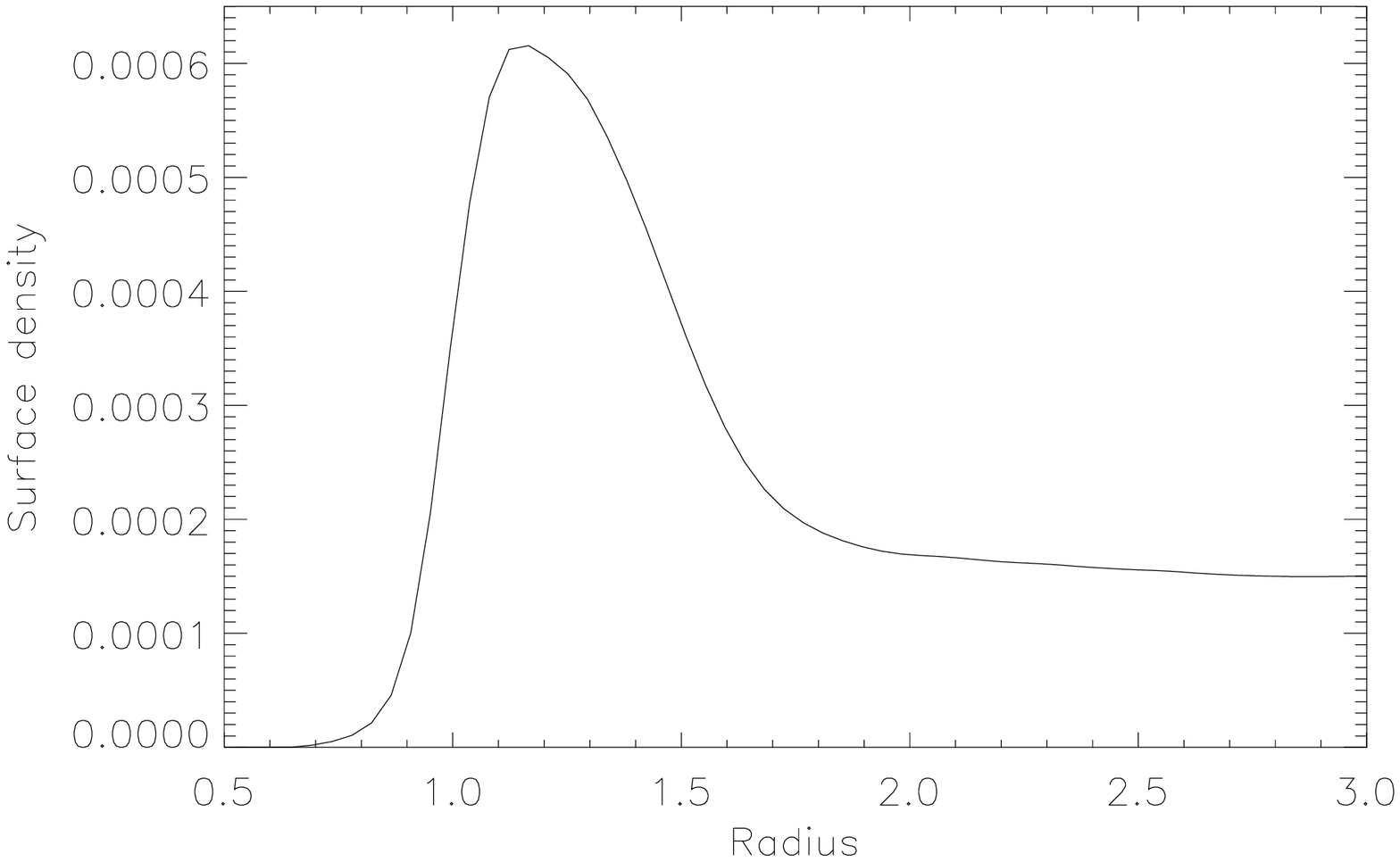} }
\resizebox{0.47\linewidth}{!}{%
\includegraphics{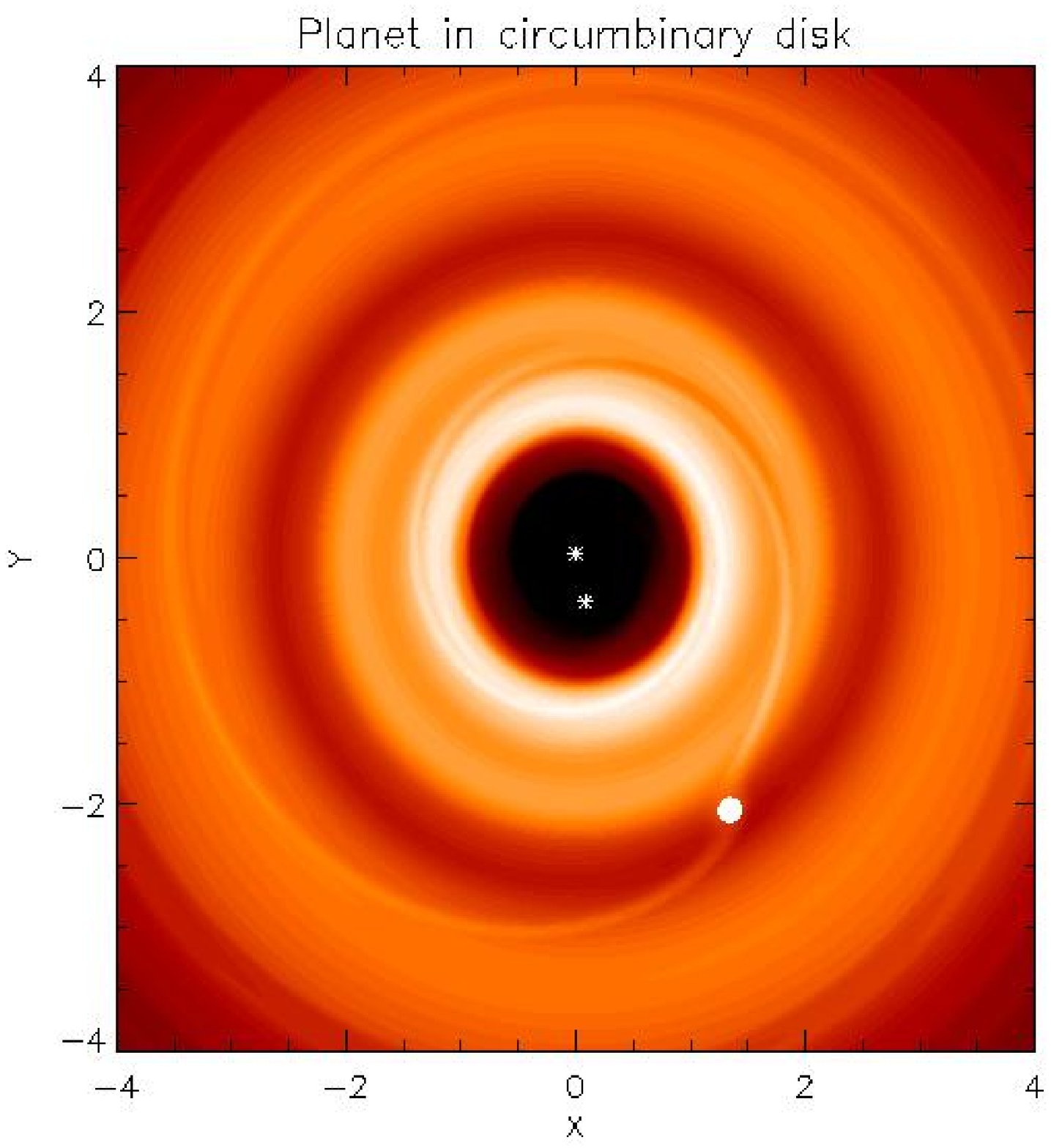} }
\caption[]
{
  The left panel shows the azimuthally averaged surface density
  profile at the end of the simulation. The right panel shows
  an image of the disk along with the protoplanet and binary system.
  This image corresponds to an earlier time during which the planet is
  migrating inward toward the central binary system.
}
\label{circbin3}
\end{figure}

Once the binary eccentricity
reaches a constant value, a low mass protoplanet
($m_p=50$ M$_{\oplus}$) was inserted in the disk 
on a circular orbit with semimajor axis $a_p=3$ AU
and allowed to evolve. The planet migrates
inward due to interaction with the disk, as shown
in figure~\ref{circbin2}, which also shows the
planet eccentricity evolution.
As the planet semimajor axis reaches a value of $a_{bin} \simeq 1.1$ AU,
we see that migration suddenly stalls. This halting
of migration appears to be robust, and occurs for planets whose
masses that are too small for gap formation in the gas disk to occur
(Pierens \& Nelson 2007 - in preparation). We ascribe this beviour to an
increase in the corotation torque as the planet enters the inner
cavity that is cleared by the tidal torques of the binary.
A similar effect has been described by \citet{2006ApJ...642..478M}
who show that planet migration can be halted
due to the action of corotation torques at surface density
transitions. As such, we expect this stalling of migration for
low mass planets to be a generic feature within circumbinary disks,
and to occur near the edge of the tidally truncated cavity
generated by the binary. The left panel of figure~\ref{circbin3}
shows the azimuthally averaged surface density in the disk
as a function of radius at the end of the simulation, and illustrates
the point that the planet stalls within the inner cavity due to
corotation torques. The right panel shows an image of the binary,
protoplanet and circumbinary disk at the end of the simulation.

\subsection{High mass circumbinary planets}

\begin{figure}
\begin{center}
\resizebox{0.47\linewidth}{!}{%
\includegraphics{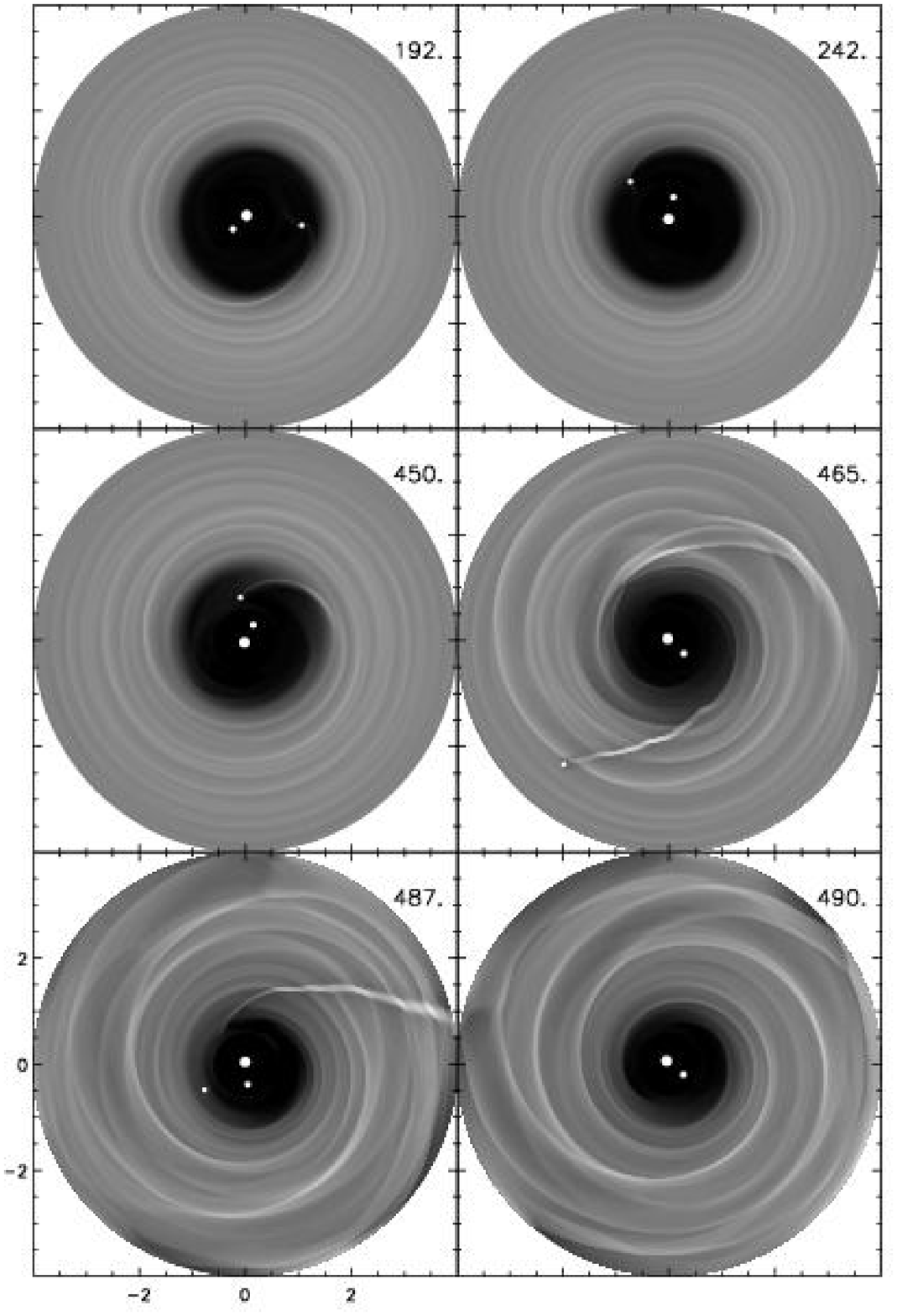} }
\caption[]
{This figure shows surface density contours for run in which the planet is ejected by the binary}
\label{circbin4}
\end{center}
\end{figure}
The simulations described below were evolved using the hydrodynamics
code {\tt NIRVANA} (Ziegler \& Yorke 1997).
The viscosity parameter $\alpha=5 \times 10^{-3}$, and the surface
density was normalised such that the disk contains about 4 Jupiter
masses interior to the initial planet semimajor axis (Nelson 2003).

The total mass of the binary plus protoplanet system is assumed to be 
1 M$_{\odot}$. We use units in which the gravitational
constant $G=1$, and the unit of length is approximately 3.6 AU.
The initial binary semimajor axis is $a_{bin}=0.4$ in our computational
units, 
and the initial planet semimajor axis $a_p=1.4$, corresponding
to 5 AU in physical units. Thus the planet lies just outside
the 6:1 mean motion resonance with the binary.
Simulations were performed for a variety of initial binary eccentricities,
$e_{bin}$, and the protoplanet was
initially in circular orbit. The binary mass ratio $q_{bin}=0.1$ 
for all simulations presented in this section, but larger values
were considered in Nelson (2003).
The unit of time quoted in the discussion 
below is the orbital period at $R=1$.

\begin{figure}
\resizebox{0.47\linewidth}{!}{%
\includegraphics{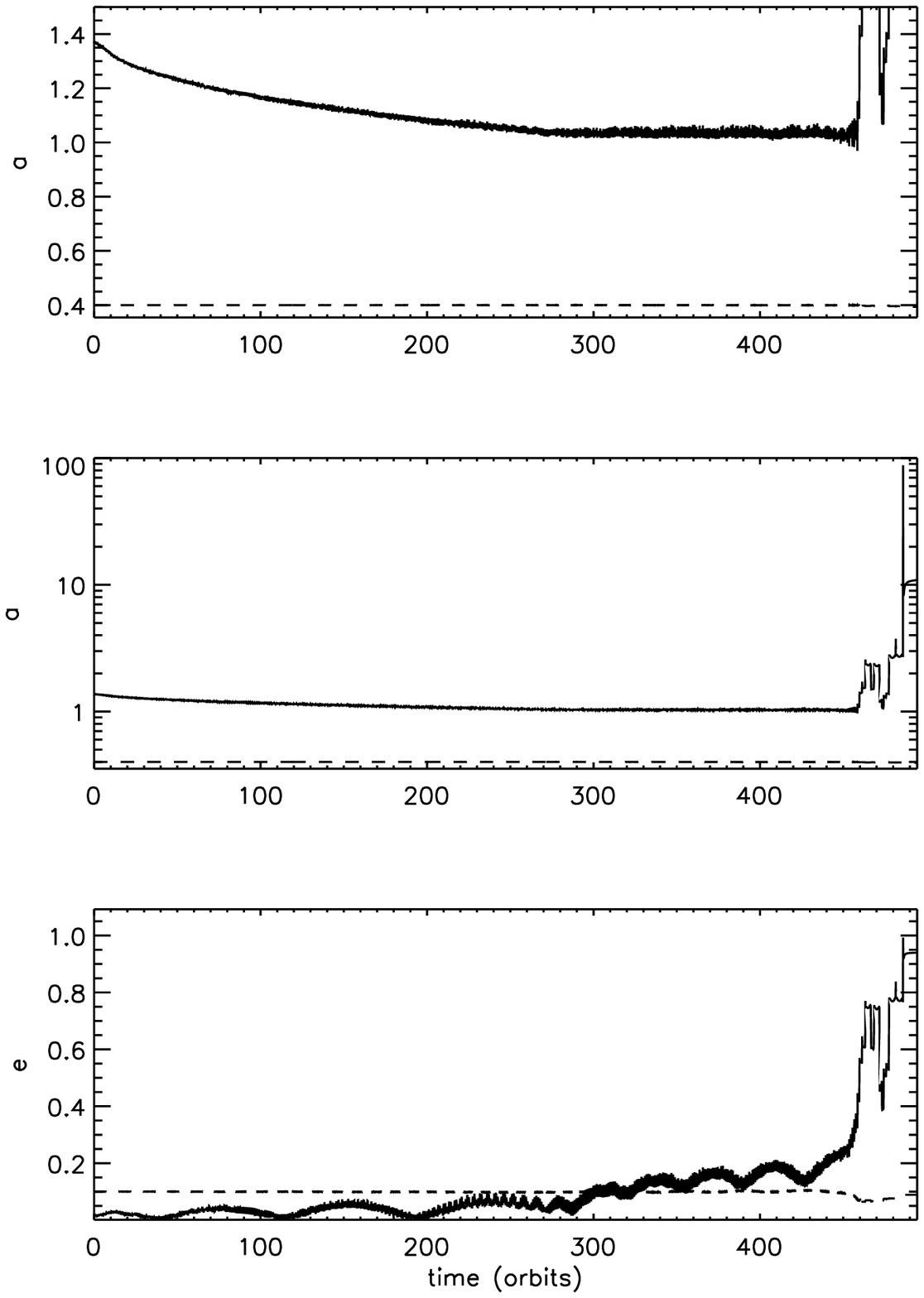} }
\resizebox{0.47\linewidth}{!}{%
\includegraphics{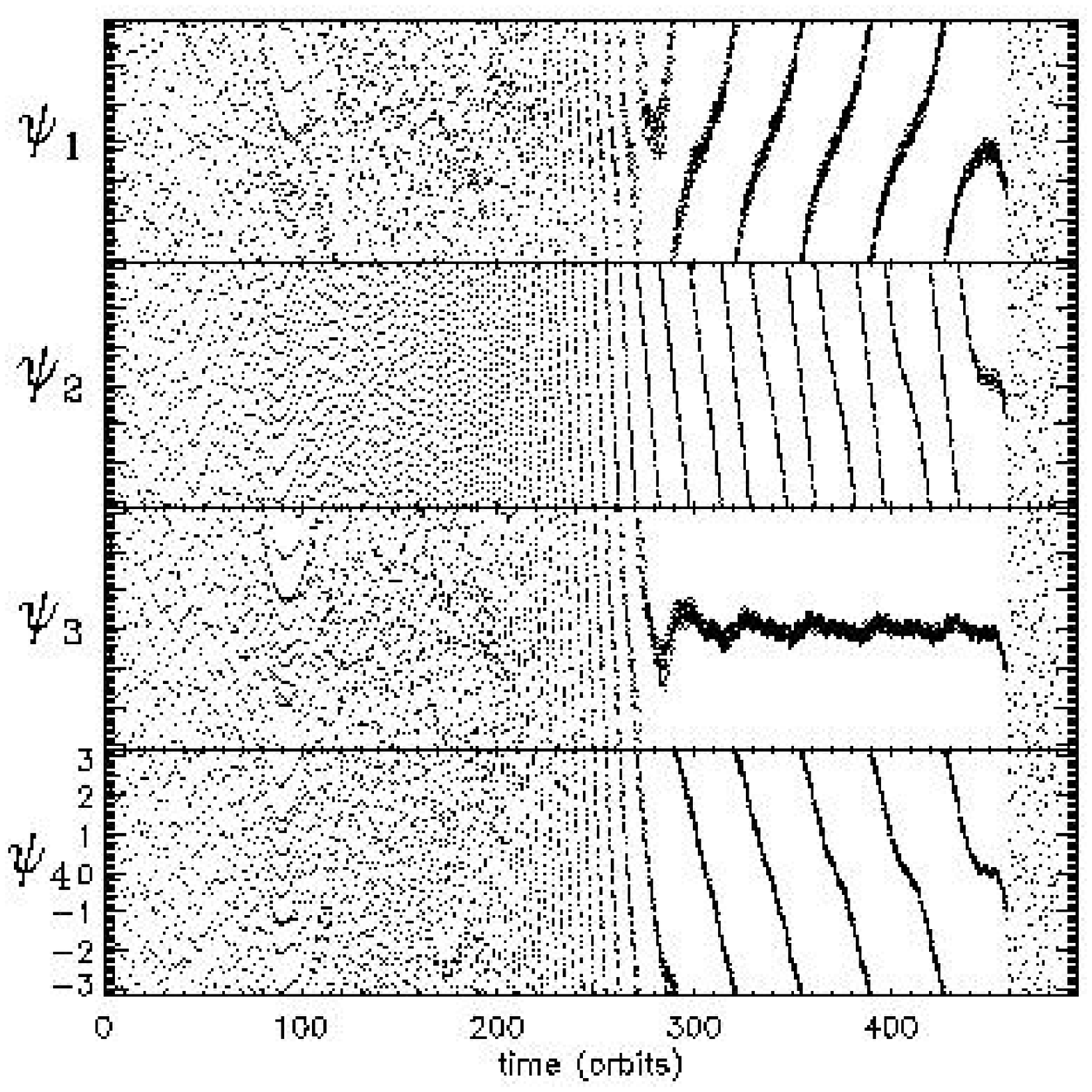} }
\caption[]
{The left panel shows the semimajor axes and eccentricities for a
run in which planet is scattered by the binary. The right panel
shows the resonant angles for the 4:1 resonance, indicating
capture into this resonance prior to scattering.}
\label{circbin5}
\end{figure}
The results of the simulations
can be divided into three main categories 
(Mode 1, Mode 2, and Mode 3), which are
described below, and are
most strongly correlated with
changes in the binary mass ratio, $q_{bin}$,  and
binary eccentricity $e_{bin}$. Changes to the disk mass and/or protoplanet
mass appear to be less important. Here we present the results of just
three simulations that illustrate these basic modes of evolution.

In some runs the planet entered the
4:1 mean motion resonance with the binary.
The associated resonant angles in the coplanar case are defined by:
\begin{eqnarray}
\psi_1 =  4 \lambda_s - \lambda_p - 3 \omega_s \;\;\;\;\;\;\;\; &&
\psi_2  = 4 \lambda_s - \lambda_p - 3 \omega_p  \\
\psi_3 =  4 \lambda_s - \lambda_p - 2 \omega_s - \omega_p &&
\psi_4  =  4 \lambda_s - \lambda_p - 2 \omega_p -\omega_s \nonumber
\label{res_ang}
\end{eqnarray}
where $\lambda_s$, $\lambda_p$ are the mean longitudes of the secondary star
and protoplanet, respectively, and $\omega_s$, $\omega_p$ are the longitudes
of pericentre of the secondary and protoplanet, respectively. When in
resonance $\psi_3$ or $\psi_4$ should librate,
or all the angles should librate.
In principle the protoplanet is able to enter higher order resonances
than 4:1, such as 5:1 or 6:1, since its initial location lies
beyond these resonance locations. However, none of the simulations presented
here resulted in such a capture. Test calculations
indicate that capture into higher order resonances requires slower planetary
migration rates than those that arise in these simulations. For significantly
faster migration rates the planet may pass through the 4:1 resonance
(Nelson 2003).

\subsubsection{Mode 1 -- Planetary Scattering}
\label{sec:mode1}
A number of simulations resulted
in a close encounter between the protoplanet and binary system, leading to
gravitational scattering of the protoplanet to larger radii, or into an
unbound state. We label this mode of evolution as `Mode 1'.
Typically the initial scattering
causes the eccentricity of the planet to grow to values
$e_p \simeq 0.9$,
and the semimajor axis to increase to $a_p \simeq 6$ -- 8. In runs that
were continued for significant times after this initial scattering,
ejection of the planet could occur after subsequent
close encounters.

We illustrate this mode of evolution using a simulation with
$m_p= 3$ Jupiter masses and $q_{bin}=0.1$. A series of snapshots of
the simulation are shown in figure~\ref{circbin4}.
Mode 1 evolution proceeds as follows.
The protoplanet migrates in toward the
central binary due to interaction with the circumbinary disk, and
temporarily enters the 4:1 mean motion resonance with the binary.
The migration and eccentricity evolution is shown in the left panel of
figure~\ref{circbin5}, and the resonance angles are shown in the right panel.
The resonant angle $\psi_3$ librates with low amplitude,
indicating that the protoplanet is strongly locked in the resonance.
The resonance drives the eccentricity of the protoplanet upward, until
the protoplanet has a close encounter with the secondary star during or close
to periapse, and is scattered out of the resonance into a 
high eccentricity orbit with significantly larger semimajor axis.
We note that the existence of a resonance normally helps maintain the
stability of two objects orbiting about a central mass. However, when
one of the objects is a star, the large perturbations experienced by the
planet can cause the resonance to break when the eccentricities are 
significant. Once out of resonance, the chances of a close encounter and
subsequent scattering are greatly increased. This provides a method
of forming `free--floating planets'.

\subsubsection{Mode 2 -- Near--resonant Protoplanet}
\label{sec:mode2}

\begin{figure}
\begin{center}
\resizebox{0.47\linewidth}{!}{%
\includegraphics{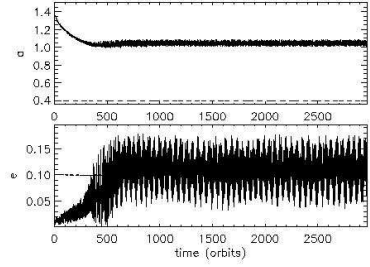} }
\caption[]
{This figure shows semimajor axes and eccentricities for the Mode 2 run
described in the text.}
\label{circbin6}
\end{center}
\end{figure}

A mode of evolution was found in some of the simulations leading
to the protoplanet orbiting stably just outside of the 4:1
resonance. We label this mode of evolution as `Mode 2'.
Mode 2 evolution is illustrated by a simulation for which $m_p=1$,
$q_{bin}=0.1$, and $e_{bin}=0.1$. The evolution of the orbital elements
are shown in figure~\ref{circbin6}.
Here, the protoplanet migrates inward and
becomes {\em weakly} locked into the 4:1 resonance,
with the resonant angle $\psi_3$ librating with large
amplitude. The resonance becomes undefined and breaks when
$e_p=0$ momentarily during the high amplitude oscillations of $e_p$
that accompany the libration of $\psi_3$.
The protoplanet undergoes a period
of outward migration through interaction with the disk
by virtue of the eccentricity having attained values of 
$e_p \simeq 0.17$ once the resonance is broken.
Unpublished simulations show that gap--forming protoplanets
orbiting in tidally truncated disks undergo outward migration if they
are given eccentricities of this magnitude impulsively.
The outward migration moves the planet to a safer
distance away from the binary, helping to avoid instability.  \\
Once the protoplanet has migrated to just beyond the 4:1 resonance the
outward migration halts, since its eccentricity reduces
slightly, and the planet remains there for the duration of the simulation.
The system achieves a balance between eccentricity damping by the disk and
eccentricity excitation by the binary, maintaining a mean value of 
$e_p \simeq 0.12$ (Nelson 2003). The torque exerted by the disk on the
protoplanet is significantly weakened by virtue of the finite
eccentricity (Nelson 2003), preventing the planet from migrating back
toward the binary. \\
Continuation of this run in the absence of the
disk indicates that the planet remains stable for over $6 \times 10^6$ orbits.
This is in good agreement with the stability criteria obtained by
Holman \& Wiegert (1999) since the protoplanet lies just 
outside of the zone of instability found by their study.
\subsubsection{Mode 3 -- Eccentric Disk}
\label{sec:mode3}
\begin{figure}
\resizebox{0.47\linewidth}{!}{%
\includegraphics{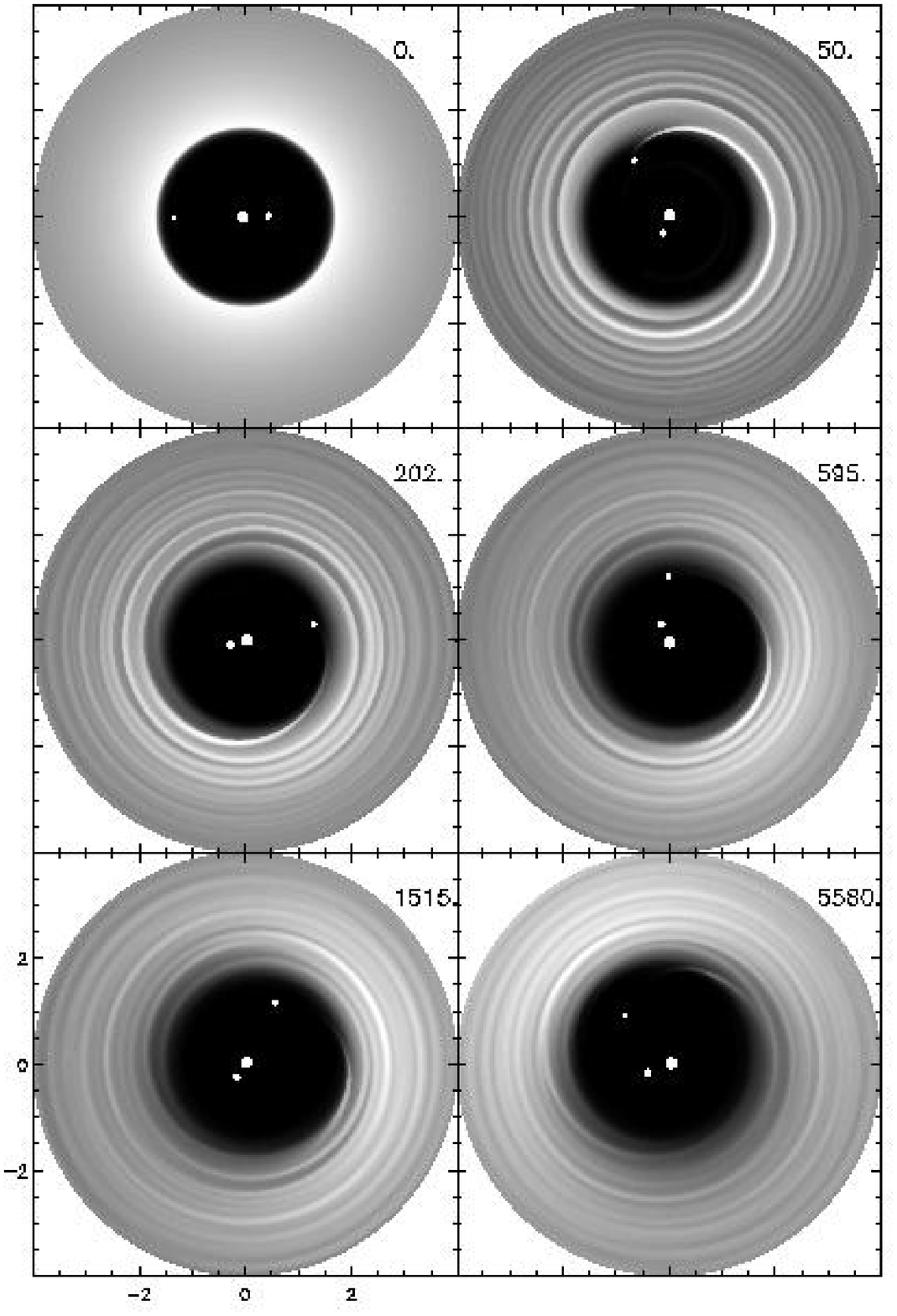} }
\resizebox{0.47\linewidth}{!}{%
\includegraphics{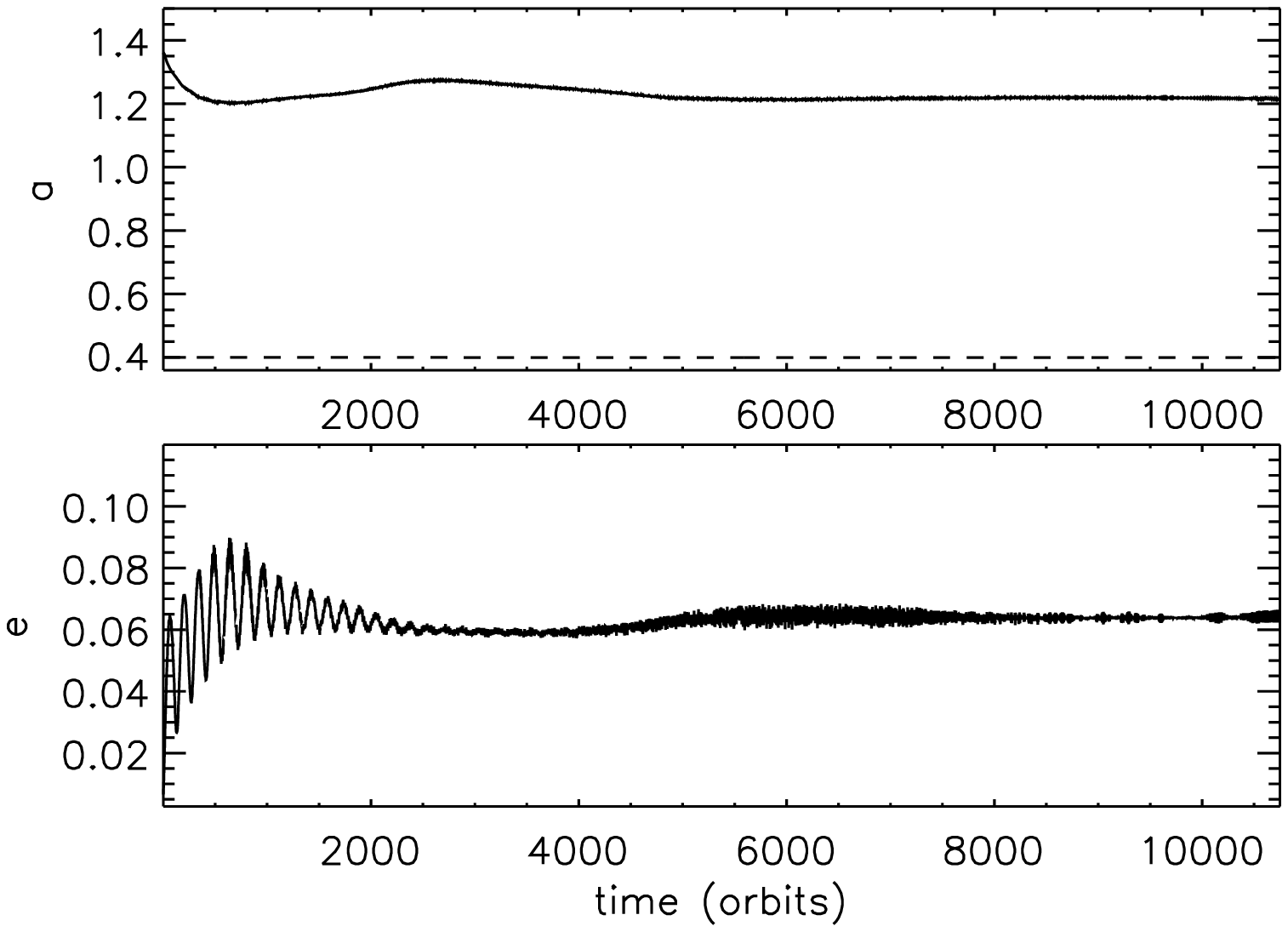} }
\caption[]
{The left panel shows contours of surface density for the Mode 3 run described
in the text. The right panel shows the resulting changes to the 
semimajor axis and eccentricity of the protoplanet.}
\label{circbin7}
\end{figure}
A mode of evolution was found in which the planetary migration
was halted before the protoplanet could approach the central binary
and reach the 4:1 resonance. This only occurred when the central binary
had an initial eccentricity of $e_{bin} \ge 0.2$. The
migration stalls because the circumbinary disk becomes eccentric.
We label this mode of evolution as `Mode 3', and illustrate it
using a simulation
with $m_p=1$ Jupiter mass, $q_{bin}=0.1$, and $e_{bin}=0.2$.
The left panel of figure~\ref{circbin7} shows snapshots of the 
surface density at different times
during the simulation, with the disk becoming noticeably eccentric.
Interaction between the protoplanet and the eccentric disk leads to 
a dramatic reduction or even reversal of the time--averaged
torque driving the migration.
This is because the disk--planet interaction becomes dominated by the $m=1$
surface density perturbation in the disk rather than by the usual interaction
at Lindblad resonances in the disk. Linear calculations of planets 
orbiting in eccntric disks also show the possibility of outward
or stalled migration \citep{2002A&A...388..615P}.

The right panel of figure~\ref{circbin7} shows the 
evolution of the semimajor axis
and eccentricity of the planet, illustrating the stalled migration.
Simulations of this type can be run for many thousands of planetary
orbits without any significant net inward migration occurring.
Such systems are likely to be stable long after the circumbinary
disk has dispersed, since the planets remain in the region of stability 
defined by the work of \citet{1999AJ....117..621H} and are 
probably the best candidates for finding stable circumbinary extrasolar
planets. Interestingly, spectroscopic binary systems with significant
eccentricity are significantly more
numerous than those with lower eccentricities
\citep{1991A&A...248..485D, 2000prpl.conf..703M}, suggesting
that circumbinary planets may be common if planets are able to form in
circumbinary disks.

\section{Conclusions} 
Much of the work presented in this article is preliminary, and so the
following statements should be viewed with the necessary caution.
The conclusions about planet formation and evolution in binary systems
that we are able to draw thus far are:
\begin{itemize}
\item{In systems such as $\gamma$ Cep, the nascent circumstellar disk
is expected to be tidally truncated at a radius of $\simeq 4$ AU, and
to be driven into an eccentric and precessing state by the
binary gravitational potential}
\item{A low mass planet that forms in such a disk will itself become
eccentric, and will migrate inward on a fairly rapid time scale}
\item{Gas accretion onto such a planet is likely to be highly efficient
because of the induced orbital eccentricity, such that a large fraction
of the disk gas will accrete onto the planet. Simulations indicate
that a gas disk containing $\simeq 3$ Jupiter masses will form a
planet of $\simeq 2$ Jupiter masses, as required to fit the minimum mass
of the planet detected in the $\gamma$ Cep system.}
\item{Simulations of planetesimals orbiting in a tidal truncated
and eccentric protoplanetary disk indicate that high velocity collisions
are likely. Such collisions will probably lead to fragmentation of the
planetesimals rather than their growth. Further work is
required to confirm this picture.}
\item{Low mass planets in circumbinary disk migrate inward
until they reach the gap edge, where they appear to stall
due to the action of corotation torques.}
\item{Should these low mass planets grow to become gas giants,
a range of outcomes seem likely. These include stalled migration
leading to the formatuon of stable circumbinary giant planets, and
inward migration followed by scattering and ejection by the central binary.}
\end{itemize}

%
\bibliographystyle{aa}
\bibliography{kley8,kleyx}

\begin{thebibliography}{35}
\expandafter\ifx\csname natexlab\endcsname\relax\def\natexlab#1{#1}\fi

\bibitem[{{Armitage} {et~al.}(1999){Armitage}, {Clarke}, \&
  {Tout}}]{1999MNRAS.304..425A}
{Armitage}, P.~J., {Clarke}, C.~J., \& {Tout}, C.~A. 1999, \mnras, 304, 425

\bibitem[{{Artymowicz} \& {Lubow}(1994)}]{1994ApJ...421..651A}
{Artymowicz}, P. \& {Lubow}, S.~H. 1994, \apj, 421, 651

\bibitem[{{D'Angelo} {et~al.}(2003){D'Angelo}, {Kley}, \&
  {Henning}}]{2003ApJ...586..540D}
{D'Angelo}, G., {Kley}, W., \& {Henning}, T. 2003, \apj, 586, 540

\bibitem[{{de Val-Borro} {et~al.}(2006){de Val-Borro}, {Edgar}, {Artymowicz},
  {Ciecielag}, {Cresswell}, {D'Angelo}, {Delgado-Donate}, {Dirksen}, {Fromang},
  {Gawryszczak}, {Klahr}, {Kley}, {Lyra}, {Masset}, {Mellema}, {Nelson},
  {Paardekooper}, {Peplinski}, {Pierens}, {Plewa}, {Rice}, {Sch{\"a}fer}, \&
  {Speith}}]{2006MNRAS.370..529D}
{de Val-Borro}, M., {Edgar}, R.~G., {Artymowicz}, P., {et~al.} 2006, \mnras,
  370, 529

\bibitem[{{Duquennoy} \& {Mayor}(1991)}]{1991A&A...248..485D}
{Duquennoy}, A. \& {Mayor}, M. 1991, \aap, 248, 485

\bibitem[{{Eggenberger} {et~al.}(2004){Eggenberger}, {Udry}, \&
  {Mayor}}]{2004A&A...417..353E}
{Eggenberger}, A., {Udry}, S., \& {Mayor}, M. 2004, \aap, 417, 353

\bibitem[{{Hatzes} {et~al.}(2003){Hatzes}, {Cochran}, {Endl}, {McArthur},
  {Paulson}, {Walker}, {Campbell}, \& {Yang}}]{2003ApJ...599.1383H}
{Hatzes}, A.~P., {Cochran}, W.~D., {Endl}, M., {et~al.} 2003, \apj, 599, 1383

\bibitem[{{Holman} \& {Wiegert}(1999)}]{1999AJ....117..621H}
{Holman}, M.~J. \& {Wiegert}, P.~A. 1999, \aj, 117, 621

\bibitem[{{Innanen} {et~al.}(1997){Innanen}, {Zheng}, {Mikkola}, \&
  {Valtonen}}]{1997AJ....113.1915I}
{Innanen}, K.~A., {Zheng}, J.~Q., {Mikkola}, S., \& {Valtonen}, M.~J. 1997,
  \aj, 113, 1915

\bibitem[{{Kley}(1989)}]{1989A&A...208...98K}
{Kley}, W. 1989, \aap, 208, 98

\bibitem[{{Kley}(1999)}]{1999MNRAS.303..696K}
{Kley}, W. 1999, \mnras, 303, 696

\bibitem[{{Kley}(2000)}]{2000IAUS..200P.211K}
{Kley}, W. 2000, in IAU Symposium, 211P

\bibitem[{{Kley} \& {Burkert}(2000)}]{2000ASPC..219..189K}
{Kley}, W. \& {Burkert}, A. 2000, in ASP Conf. Ser. 219: Disks, Planetesimals,
  and Planets, 189

\bibitem[{{Larwood} {et~al.}(1996){Larwood}, {Nelson}, {Papaloizou}, \&
  {Terquem}}]{1996MNRAS.282..597L}
{Larwood}, J.~D., {Nelson}, R.~P., {Papaloizou}, J.~C.~B., \& {Terquem}, C.
  1996, \mnras, 282, 597

\bibitem[{{Lissauer} {et~al.}(2004){Lissauer}, {Quintana}, {Chambers},
  {Duncan}, \& {Adams}}]{2004RMxAC..22...99L}
{Lissauer}, J.~J., {Quintana}, E.~V., {Chambers}, J.~E., {Duncan}, M.~J., \&
  {Adams}, F.~C. 2004, in Revista Mexicana de Astronomia y Astrofisica
  Conference Series, 99--103

\bibitem[{{Masset} {et~al.}(2006){Masset}, {Morbidelli}, {Crida}, \&
  {Ferreira}}]{2006ApJ...642..478M}
{Masset}, F.~S., {Morbidelli}, A., {Crida}, A., \& {Ferreira}, J. 2006, \apj,
  642, 478

\bibitem[{{Mathieu} {et~al.}(2000){Mathieu}, {Ghez}, {Jensen}, \&
  {Simon}}]{2000prpl.conf..703M}
{Mathieu}, R.~D., {Ghez}, A.~M., {Jensen}, E.~L.~N., \& {Simon}, M. 2000,
  Protostars and Planets IV, 703

\bibitem[{{Monin} {et~al.}(2007){Monin}, {Clarke}, {Prato}, \&
  {McCabe}}]{2007prpl.conf..395M}
{Monin}, J.-L., {Clarke}, C.~J., {Prato}, L., \& {McCabe}, C. 2007, in
  Protostars and Planets V, ed. B.~{Reipurth}, D.~{Jewitt}, \& K.~{Keil},
  395--409

\bibitem[{{Mugrauer} {et~al.}(2007){Mugrauer}, {Neuhaeuser}, \&
  {Mazeh}}]{2007astro.ph..3795M}
{Mugrauer}, M., {Neuhaeuser}, R., \& {Mazeh}, T. 2007, ArXiv Astrophysics
  e-prints

\bibitem[{{Mugrauer} \& {Neuh{\"a}user}(2005)}]{2005MNRAS.361L..15M}
{Mugrauer}, M. \& {Neuh{\"a}user}, R. 2005, \mnras, 361, L15

\bibitem[{{Nelson}(2000)}]{2000ApJ...537L..65N}
{Nelson}, A.~F. 2000, \apjl, 537, L65

\bibitem[{{Nelson}(2003)}]{2003MNRAS.345..233N}
{Nelson}, R.~P. 2003, \mnras, 345, 233

\bibitem[{{Nelson} {et~al.}(2000){Nelson}, {Papaloizou}, {Masset}, \&
  {Kley}}]{2000MNRAS.318...18N}
{Nelson}, R.~P., {Papaloizou}, J.~C.~B., {Masset}, F.~S., \& {Kley}, W. 2000,
  \mnras, 318, 18

\bibitem[{{Neuh{\"a}user} {et~al.}(2007){Neuh{\"a}user}, {Mugrauer},
  {Fukagawa}, {Torres}, \& {Schmidt}}]{2007A&A...462..777N}
{Neuh{\"a}user}, R., {Mugrauer}, M., {Fukagawa}, M., {Torres}, G., \&
  {Schmidt}, T. 2007, \aap, 462, 777

\bibitem[{{Papaloizou}(2002)}]{2002A&A...388..615P}
{Papaloizou}, J.~C.~B. 2002, \aap, 388, 615

\bibitem[{{Papaloizou}(2005)}]{2005A&A...432..757P}
{Papaloizou}, J.~C.~B. 2005, \aap, 432, 757

\bibitem[{{Pierens} {et~al.}(2005){Pierens}, {Dutrey}, {Guilloteau}, \&
  {Hur{\'e}}}]{2005sf2a.conf..733P}
{Pierens}, A., {Dutrey}, A., {Guilloteau}, S., \& {Hur{\'e}}, J.-M. 2005, in
  SF2A-2005: Semaine de l'Astrophysique Francaise, ed. F.~{Casoli},
  T.~{Contini}, J.~M. {Hameury}, \& L.~{Pagani}, 733--+

\bibitem[{{Pierens} \& {Nelson}(2007)}]{2007MNRAS..submitted}
{Pierens}, A. \& {Nelson}, R. 2007, \mnras, 00, 00

\bibitem[{{Queloz} {et~al.}(2000){Queloz}, {Mayor}, {Weber}, {Bl{\'e}cha},
  {Burnet}, {Confino}, {Naef}, {Pepe}, {Santos}, \&
  {Udry}}]{2000A&A...354...99Q}
{Queloz}, D., {Mayor}, M., {Weber}, L., {et~al.} 2000, \aap, 354, 99

\bibitem[{{Quintana} {et~al.}(2007){Quintana}, {Adams}, {Lissauer}, \&
  {Chambers}}]{2007astro.ph..1266Q}
{Quintana}, E.~V., {Adams}, F.~C., {Lissauer}, J.~J., \& {Chambers}, J.~E.
  2007, ArXiv Astrophysics e-prints

\bibitem[{{Th{\'e}bault} {et~al.}(2004){Th{\'e}bault}, {Marzari}, {Scholl},
  {Turrini}, \& {Barbieri}}]{2004A&A...427.1097T}
{Th{\'e}bault}, P., {Marzari}, F., {Scholl}, H., {Turrini}, D., \& {Barbieri},
  M. 2004, \aap, 427, 1097

\bibitem[{{Torres}(2007)}]{2007ApJ...654.1095T}
{Torres}, G. 2007, \apj, 654, 1095

\bibitem[{{Turrini} {et~al.}(2005){Turrini}, {Barbieri}, {Marzari}, {Thebault},
  \& {Tricarico}}]{2005MSAIS...6..172T}
{Turrini}, D., {Barbieri}, M., {Marzari}, F., {Thebault}, P., \& {Tricarico},
  P. 2005, Memorie della Societa Astronomica Italiana Supplement, 6, 172

\bibitem[{{Weidenschilling}(1977)}]{1977MNRAS.180...57W}
{Weidenschilling}, S.~J. 1977, \mnras, 180, 57

\bibitem[{{Ziegler} \& {Yorke}(1997)}]{1997ZiegYork}
{Ziegler}, U. \& {Yorke}, H. 1997, Computer Physics Communications, 101, 54

\end{thebibliography}
%


\printindex
\end{document}